\documentclass[
preprint,amsmath,superscriptaddress,prl]{revtex4}  

\usepackage[dvips]{graphicx}
\usepackage{epsfig}
\usepackage{color}
\usepackage{subfigure}
\usepackage{soul}
\usepackage{enumerate}

\begin{document} 
\title{
Lognormals, Power Laws and Double Power Laws
in the Distribution of Frequencies
of Harmonic Codewords from 
Classical Music
} 
\author{Marc Serra-Peralta}
\affiliation{%
Centre de Recerca Matem\`atica,
Edifici C, Campus Bellaterra,
E-08193 Barcelona, Spain
}
\affiliation{Departament de Física,
Facultat de Ci\`encies,
Universitat Aut\`onoma de Barcelona,
E-08193 Barcelona, Spain
}
\author{Joan Serr\`a}
\affiliation{Dolby Laboratories, E-08018 Barcelona, Spain}
\author{\'Alvaro Corral}
\affiliation{%
Centre de Recerca Matem\`atica,
Edifici C, Campus Bellaterra,
E-08193 Barcelona, Spain
}\affiliation{Departament de Matem\`atiques,
Facultat de Ci\`encies,
Universitat Aut\`onoma de Barcelona,
E-08193 Barcelona, Spain}
\affiliation{Complexity Science Hub Vienna,
Josefst\"adter Stra$\beta$e 39,
1080 Vienna,
Austria
}

\begin{abstract} 
Zipf's law is a paradigm describing the importance of different elements
in communication systems, 
especially in linguistics.
Despite the complexity of the hierarchical structure of language, 
music has in some sense an even more complex structure,
due to its multidimensional character
(melody, harmony, rhythm, timbre, etc.).
Thus, the relevance of Zipf's law in music is still an open question.
Using discrete codewords representing harmonic content
obtained from a large-scale analysis of classical composers,
we show that a nearly universal Zipf-like law holds at a qualitative level.
{However, in an in-depth quantitative analysis, 
where we introduce the double power-law distribution
as a new player
in the classical debate between the superiority of Zipf's (power) law
and that of the lognormal distribution,
we conclude 
not only that universality does not hold, 
but also that there is not a unique probability distribution that best describes
the usage of the different codewords by each composer.}
%
%
\end{abstract} 


\maketitle

\section{Introduction}

For centuries, physics has dealt with deterministic mathematical laws, 
such as Newton's laws of mechanics, the laws of electromagnetism, or the laws of relativity~\cite{longair_2003}.
It were Maxwell and Boltzmann who, in the 19th century, 
discovered probabilistic or statistical laws in the study of the motion of the (hypothetical) 
particles constituting a gas.
The so-famous Planck's radiation law 
can be understood as another instance of a probabilistic law.
The great insight of Maxwell, Boltzmann, and Planck (and others, like Einstein)
was the introduction of probability
to infer the mechanics of the constituents
of matter and radiation.
That insight has been one of the most successful 
knowledge programs
in the history of humankind.
Although the just-mentioned examples of probabilistic laws are valid in the ``ideal case''
(no interaction between the constituents),
an interaction (at least with the surroundings) 
has to be present to ensure the existence of a state of equilibrium.
In general, 
the study of how macroscopic behavior emerges from microscopic interactions
(either in equilibrium or out of equilibrium)
is the goal of statistical physics.


In ecology, in the social sciences (sociology, economics, demography), 
and in the study of 
technological and information networks,
one is, in some sense, in a situation similar to statistical physics, 
in which there is an enormous number of individual entities 
whose behavior 
depends on each other,
leading to an emerging collective behavior
\cite{Watts,West_book}.
Despite the lack of well-defined underlying microscopic laws,
it is remarkable that one may find 
regular statistical laws describing the aggregated properties of the constituent entities,
and even more remarkable that
it is the same law which seems to capture
a particular but important aspect of many of these systems
\cite{Camacho_sole,Axtell,Adamic_Huberman,%
Pueyo}.
This ubiquitous and nearly universal law is Zipf's law~\cite{Li02}, 
which describes how the constituent entities, or tokens, 
are distributed into larger groups, or types. 
In this way, Zipf's law states that the size distribution of these groups
(measured in terms of constituent entities) 
follows a power-law distribution with a loosely constrained value of the exponent, 
close to 2 (for the probability mass function).
In an additional twist in favor of the meaningfulness of statistical ``natural'' laws beyond physics, 
Zipf's law emerges again
when one considers quantitative approaches to human sciences 
(mainly the study of language \cite{Baayen,Baroni2009,Zanette_book,Piantadosi,Moreno_Sanchez}),
where the nature of the interactions between the constituent entities is not so clear \cite{Stephens_Bialek,Corral_muro}.

However, there have been serious issues with the Zipf's paradigm, 
and with power laws in general,
being the most important of them 
the lack of generality of the results \cite{Moreno_Sanchez,Gerlach_Font_Clos}
and
low statistical rigor \cite{Bauke,White,Clauset,Corral_Deluca,Gerlach_Altmann_prl,Corral_Gonzalez}.
In the first case, 
a very small number of datasets are usually analyzed 
in order to establish
the validity of Zipf's law in every particular system.
For instance, in quantitative linguistics, research articles are usually focused in about a dozen 
(or even less) texts \cite{Font-Clos2013,Corral_Boleda}, 
with the selection of them 
seeming rather arbitrary.
Therefore, many published claims should be considered as anecdotic examples, 
or conjectures, rather than well-established facts.

Regarding statistical rigor,
proper fitting methods and goodness-of-fit tests have seldomly been used, 
being replaced many times by visual, qualitative checks. 
Moreover, some apparently rigorous procedures \cite{Clauset} 
have been found to yield inconsistent results \cite{Corral_nuclear,Voitalov_krioukov}.
Therefore, it is not yet established for which systems 
Zipf's law is a rough or even bad approximation
and for which systems it is a good description in some range or limit.
{An annoying side effect of the lack of proper statistical tools 
is the recurrent debate about
if the lognormal distribution is superior or not to the power law 
to describe (some) Zipfian systems \cite{Malevergne_Sornette_umpu,Corral_Arcaute}.}
A further concern is some ambiguity in the definition of Zipf's law
\cite{Mandelbrot61,Moreno_Sanchez,Corral_Cancho}, 
which admits several mathematical formulations 
not strictly equivalent between them.

In recent years, several authors have tried to overcome the problems of Zipf's law in linguistics.
For instance, Moreno-S\'anchez et al.~\cite{Moreno_Sanchez}
analyzed different mathematical alternatives to Zipf's law, 
using all English texts (tens of thousands) 
available from the Project Gutenberg digital library.
In a sort of complementary study,
Mehri and Jamaati \cite{Mehri}
considered just one text (the Bible)
but in its translation to one hundred distinct languages.
An alternative 
approach, 
instead of analyzing many different individual texts, 
has been to use big corpora 
(collections such as the British National Corpus formed by gathering many text fragments).
Although this is also a valid procedure, 
one cannot assert that results
(for instance, a claimed double power-law distribution 
comprising Zipf's law for large word frequencies 
\cite{Ferrer2001a,Montemurro01,Corral_brevity}) 
are not an artifact arising from the 
mixture of rather different sort of texts \cite{Williams_Dodds}.
In other words, the statistical properties of the British National Corpus
could be different 
if the corpus were compiled in a different way
(e.g., changing the length of the selected fragments), and very different also
to the ones of a hypothetical text of the same length from a single author.


{In the last years, diverse forms of artistic expressions
have been approached through the eyes of 
complex-systems science \cite{Perc_art}.
In this paper, we deal with music.}
Music seems to be a necessary and sufficient condition for ``humanity''~\cite{Ball_music},
in the sense that music has been present across all human societies in all times, and other animal species do not seem to have real musical capabilities. 
Thus, music is a uniquely human attribute
(needless to say, if any extraterrestrial intelligence were ever discovered,
one of the first questions to figure out would be about 
its relationship with some sort of music \cite{Voyager1,Voyager2}%
).
Even more, music is one of the human activities that attracts more public interest
(e.g., at the time of writing this, out of the 10 Twitter accounts with more followers, 
6 correspond to popular musicians or musical performers 
\cite{most_followers}%
).
Certain parallelisms between natural language and music have been noted in the literature, where music has been sometimes categorized as a ``language'' \cite{Zanette_music};
nevertheless, there is no clear notion of grammar and semantic content
in music \cite{Zanette_music,Ball_music} although there exist relations in terms of rhythm, pitch, syntax and meaning \cite{Patel}.


In any case, one can conceptualize music as a succession (in time) of some musical 
descriptors or symbols, 
which can be counted in a Zipf-like manner \cite{Zanette_nature08},
with the frequency of appearance of each different symbol playing 
the role of the size of the groups (types) in which Zipfian systems
are organized. 
A remarkable problem is that, in contrast to language \cite{Corral_Boleda}, 
the individual entities to analyze in music 
can be extraordinarily elusive to establish. 
For instance,
Manaris et al.~\cite{Manaris1} mention several possibilities
involving different combinations of pitch and duration,
as well as pitch differences,
{illustrating the multidimensional character of music}.
This, together with some technicalities to deal with musical datasets,
may explain the fact that the study of Zipf's law and other linguistic laws in music has been 
substantially limited
in comparison to natural language.
Nevertheless, some precedents are of interest.
Using a simple pitch-duration pair as a metric, 
Zanette \cite{Zanette_music} fitted a variant of Zipf's law to four classical musical pieces, 
finding a rather high power-law exponent 
(up to 4.6 for the probability mass function of frequency, 
except for a piece by Arnold Sch\"onberg).
Later, Liu et al.~\cite{Liu} did the statistics of pitch jumps for five
classical composers to find even higher power-law exponents.
In a more large-scale study,
Beltr\'an del R\'{\i}o et al.~\cite{Beltran} analyzed plain pitches 
(of which there is a maximum ``vocabulary'' of 128)
in more than 1800 MIDI files, 
containing classical music, jazz, and rock,
to fit a generalization of Zipf's law. 
This, in general, resulted in a rather high power-law exponent.
In any case, these publications did not use very high statistical standards.

In a more recent attempt,
studying popular Western music,
Serr\`a et al.~\cite{Serra_scirep}  
considered the combination of pitch classes present in short time intervals 
(harmonic and melodic chords, in some sense)
to construct discretized chromas.
Aggregating individual pieces 
for fixed lustrums of the 20th century, 
and using maximum-likelihood estimation with the necessary goodness-of-fit tests, 
they found a robust exponent for the tail close to 2.2 (in agreement with Zipf's law),
which remained stable across the different historical periods 
that were analyzed.
Distributions of timbral indicators were also explored in that work, as well as by
Haro et al. \cite{Haro}; intriguingly, the latter reference found that
Zipf's law for timbre is not only fulfilled by music but also by 
speech and natural sounds (such as rain, wind, and fire).



In this paper, we analyze classical music using its 
``crystallization''
into electronic MIDI scores,
{by means of a rather large database.}
%
One could argue that, in order to access genuine expressions of music,
audio recordings 
are preferable to MIDI scores,
due to the fact that the latter may lack the richness and nuances of interpretation
\cite{Geisel_music}
(although there are MIDI files created from the life performance of a musical piece).
Nevertheless, for our purposes, scores contain 
the essence of music,
and, in the case of music previous to the 20th century, 
{they are our best remainder of the original intention of the composer.}
Moreover,  
as to undertake statistical analysis we need to deal with discretized elements,
scores provide an objective first step in such discretization.


In the next sections,
we present the characteristics of the corpus used (Sec. 2),
describe the extraction of harmonic codewords from the MIDI files (Sec. 2 and Supplementary Information, SI),
introduce the probability distributions to fit the codewords counts 
{(power law, double power law, and lognormal, 
Sec. 3),}
explain the statistical procedure (Sec. 3 and SI),
and present the results (Sec. 4).
Naturally, we end with some conclusions.
The code used in this paper is available in Github \cite{Serra_github}.

\section{Data, Processing, and Elementary Statistics} 


\subsection{Data}

As a corpus of classical music scores we use 
the \mbox{\textit{Kunstderfuge}} database \cite{Kunst}, 
including 17,419 MIDI scores of composers from the 12th to the 20th century.
Different aspects of this musical corpus have been analyzed
elsewhere \cite{Lacasa_music,Serra_Corral_richness}. 
We perform a preliminary cleaning in which we identify and remove
traditional songs, anthems, anonymous pieces, and also MIDIs
arising from live performances, leading to a remainder of
10,523 files. 
In general, these files contain the name of the composer
and an indication of the name of the piece.
Further removing 
files that we are not able to process, 
files for which we cannot obtain the bar 
(and cannot determine therefore the temporal unit), 
files corresponding to very short pieces,
and
files corresponding to repeated pieces,
we retain 
9327 of them corresponding to 76 composers, 
ranging from Guillaume Dufay (1397--1474) to Olivier Messiaen (1908--1992). 
The complete list of composers, in chronological order, 
is provided in Table \ref{table_composers}.
Details of the data cleaning and the method of detection of repeated pieces in the corpus
are provided in the Supplementary Information.


\begin{table}[h]
\begin{center}
\caption{\label{table_composers}
Name of the 76 classical composers in the 
{\it Kunstderfuge} corpus analyzed in this paper.
The order is chronological 
(from left to right and from top to bottom, established by the average between birth and death).
The results of the best fit (in {\bf bold}), 
followed by other good fits are included.
pl $=$ (simple) power law,
dpl $=$ double power law,
ln $=$ lognormal.
Number of composers fitted by a unique distribution: 14
({\bf pl} 0;
{\bf dpl} 8;
{\bf ln} 6).
Number of composers with two good fits: 61
({\bf pl}, ln 7;
{\bf dpl}, ln 25;
{\bf ln}, pl 14;
{\bf ln}, dpl 15).
}
\smallskip 
\tiny
\begin{tabular}{|lr|lr|lr|lr|}
\hline
G. Dufay: & {\bf ln}, dpl;              & J. Desprez: & {\bf ln}, dpl;        & C. de Morales: & {\bf dpl}, ln;  & G. P. da Palestrina: & {\bf ln}, dpl \\
O. Lassus: &              {\bf ln}, dpl; &  T. L. de Victoria: &  {\bf dpl}, ln; &  W. Byrd: &            {\bf dpl}, ln; &  C. Gesualdo: &          {\bf dpl}, ln \\
J. Dowland: &             {\bf dpl}, ln; &  C. Monteverdi: &      {\bf ln}, pl; &  G. Frescobaldi: &     {\bf dpl}, ln; &  S. Scheidt: &           {\bf ln}, pl \\ 
J. J. Froberger: &        {\bf dpl}, ln; &  J. B. Lully: &        {\bf dpl}, ln; &  J.-H. d’Anglebert: &   {\bf ln}, dpl; &  D. Buxtehude: &         {\bf dpl}, ln \\
J. Pachelbel: &            {\bf   ln}; &  F. Couperin: &        {\bf   dpl}; &  D. Zipoli: &           {\bf ln}, dpl; &  A. Vivaldi: &           {\bf dpl}, ln \\
J.-F. Dandrieu: &         {\bf   dpl}; &  T. Albinoni: &        {\bf dpl}, ln; &  J. S. Bach: &         {\bf --}; &  D. Scarlatti: &         {\bf dpl}, ln \\
G. F. Händel: &           {\bf   dpl}; &  J.-P. Rameau: &       {\bf dpl}, ln; &  G. P. Telemann: &     {\bf dpl}, ln; &  J. Haydn: &             {\bf   dpl} \\ 
J. G. Albrechtsberger: &   {\bf ln}, pl; &  W. A. Mozart: &       {\bf   dpl}; &  M. Clementi: &        {\bf dpl}, ln; &  L. van Beethoven: &     {\bf   dpl} \\ 
N. Paganini: &            {\bf pl}, ln; &  F. Schubert: &         {\bf   ln}; &  J. B. Cramer: &        {\bf ln}, pl; &  F. Mendelssohn: &        {\bf ln}, dpl \\
F. Chopin: &              {\bf dpl}, ln; &  R. Schumann: &        {\bf dpl}, ln; &  H. Berlioz: &          {\bf ln}, pl; &  F. Liszt: &              {\bf ln}, dpl \\
L. M. Gottschalk: &        {\bf ln}, pl; &  C.-V. Alkan: &        {\bf   dpl}; &  C. Franck: &           {\bf   ln}; &  G. Bizet: &              {\bf ln}, pl \\ 
A. Bruckner: &             {\bf   ln}; &  M. Mússorgsky: &      {\bf dpl}, ln; &  J. Brahms: &           {\bf   ln}; &  P. I. Tchaikovsky: &     {\bf ln}, dpl \\
A. Dvořák: &               {\bf ln}, dpl; &  A. Guilmant: &        {\bf pl}, ln; &  E. Grieg: &            {\bf ln}, dpl; &  C. Saint-Saëns: &       {\bf dpl}, ln \\
I. Albéniz: &              {\bf ln}, dpl; &  G. U. Fauré: &         {\bf   ln}; &  G. Mahler: &          {\bf dpl}, ln; &  C. Debussy: &            {\bf ln}, dpl \\
L. Janáček: &              {\bf ln}, pl; &  S. Joplin: &          {\bf pl}, ln; &  A. Scriabin: &         {\bf   ln}, dpl; &  M. Reger: &             {\bf dpl}, ln \\
F. Busoni: &              {\bf dpl}, ln; &  E. Satie: &            {\bf ln}, pl; &  L. Godowsky: &        {\bf dpl}, ln; &  S. Karg-Elert: &        {\bf pl}, ln \\ 
M. Ravel: &               {\bf dpl}, ln; &  O. Respighi: &         {\bf ln}, pl; &  S. Rajmáninov: &       {\bf ln}, pl; &  A. Schoenberg: &        {\bf pl}, ln \\ 
B. Bartók: &              {\bf pl}, ln; &  N. Médtner: &          {\bf ln}, pl; &  G. Gershwin: &         {\bf   ln}, pl; &  S. Prokófiev: &          {\bf ln}, pl \\ 
Í. Stravinsky: &          {\bf pl}, ln; &  P. Hindemith: &        {\bf   ln}, dpl; &  D. Shostakóvich: &    {\bf   dpl}; &  O. Messiaen: &       {\bf dpl}, ln \\
\hline
\end{tabular}
\par
\end{center}
\end{table}

\subsection{Harmonic Codewords}

Our analysis focuses on the harmonic content of music, 
understood as the combination of pitches 
across all instruments in short time frames.
A complete summary for the obtention of the elementary units in which we decompose music is provided in the Supplementary Information 
(see also Ref. \cite{Serra_Corral_richness});
Fig. \ref{fig_cartoon} provides a simple illustration.
The different steps are:

\begin{figure}[t]
\includegraphics[width=.95\columnwidth]{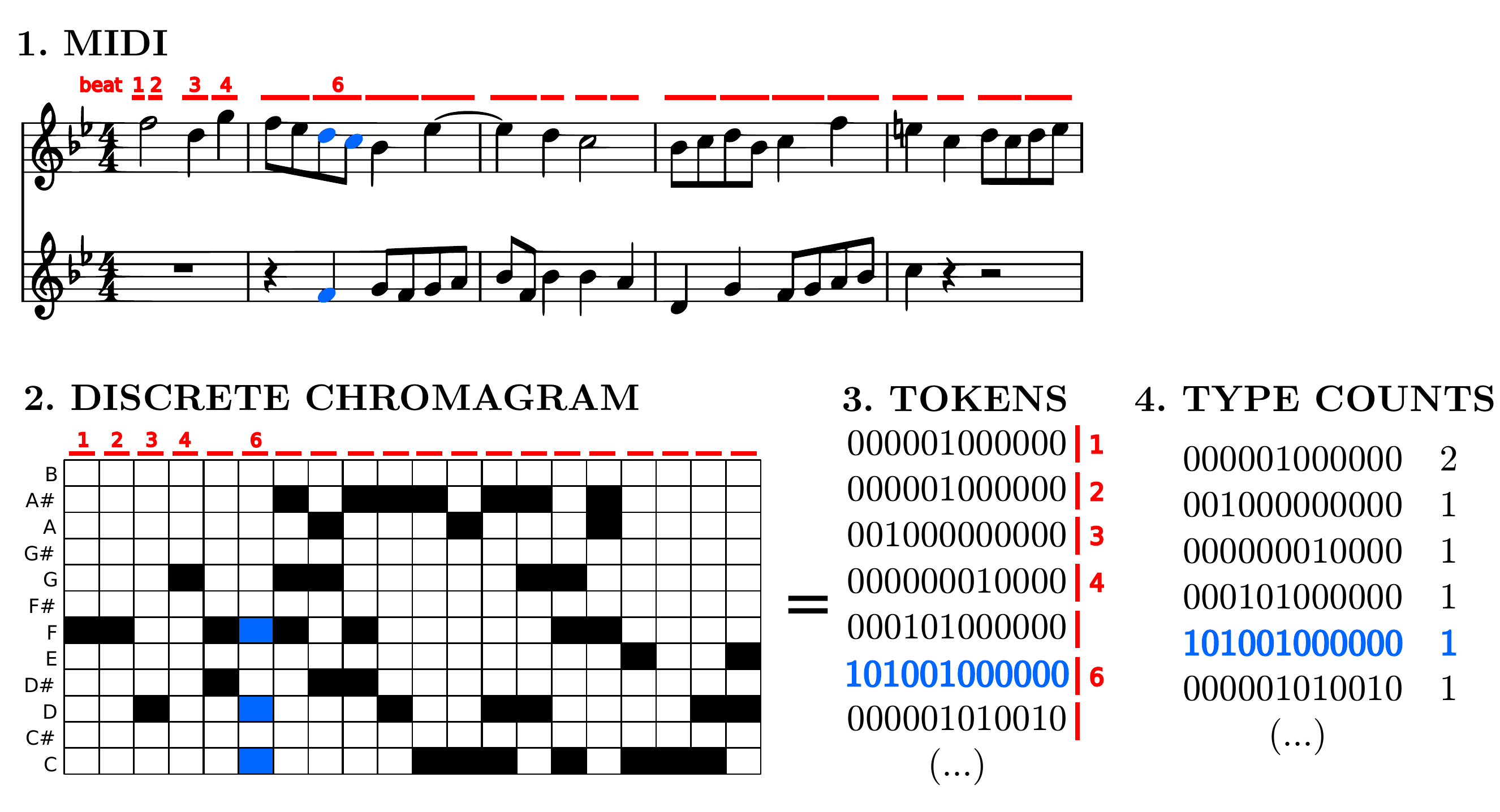}
\caption{
Scheme showing the correspondence between a score
(represented by two parallel staves)
and its representation in terms of discretized chromas.
Counts for each type (number of tokens)
are also shown.
}
\label{fig_cartoon}
\end{figure}

\begin{enumerate}[(i)]
\item Conversion of each MIDI file (corresponding to a piece)
into a text file
(containing the time occurrence, duration, and pitch of each note).

\item Transformation of pitches into (twelve) pitch classes
(i.e., collapse into a unique octave).

\item Segmentation into elementary time intervals (given by the score beats).

\item 
Construction of chromas: 12-dimensional vectors 
$(C,C\#,\dots G\#, A,A\#,B)$, 
counting the contribution of notes from each pitch class 
for each elementary time interval
and for each stave. 
That is, we collapse all staves in the piece into a single chroma sequence.

\item Discretization of chromas (using a discretization threshold),
which yields the harmonic codewords
to analyze.
These are 12-dimensional vectors of binary elements $(0,1)$.

\item Transposition to C major (major pieces)
or A minor (minor pieces).

\item For each composer, aggregation of all 
the time series of transposed discretized chromas 
(corresponding to each piece)
into a unique dataset.

\end{enumerate}

{This aggregation of pieces of the same composer 
is done in order to get significant statistics.
Although under the framework of some models explaining Zipf's law
\cite{Simon}, aggregation makes little sense, 
for other models there are no such restrictions \cite{Cattuto}.
In any case our approach is model free.
Notice that aggregation can be done in several ways.
Ours is equivalent to aggregate the frequencies $n$ of each type,
but one could also aggregate the counts $f(n)$ of each frequency $n$,
and this could be done also for relative frequencies.
The latter two options would change the results, as they lead to mixtures
of the distributions corresponding to each piece.
}

{The obtained codewords contain information about the melody 
and, mostly, the harmony of the pieces. 
The most common codewords in the studied corpus are listed in Table 6 of Ref. \cite{Serra_Corral_richness} and they are consistent with characteristic harmonic features, i.e., they correspond to harmonic sets of pitches. Therefore, the results arising from the analysis of the codewords can be associated with the harmonic characteristics of the compositions, especially in the range of high type frequencies. }

%


\subsection{Elementary Statistics}

For each dataset (corresponding to a composer), 
we count the repetitions or absolute frequency $n$ of each type 
(discretized chromas, see Fig. \ref{fig_cartoon}). 
This absolute frequency is our random variable, 
and the number of appearances of each value of the frequency
(frequencies of frequencies, then)
constitutes an empirical estimation of the probability mass function of the frequency,
which we may denote as $f(n)$.
However, due to the broad range of the distributions (with $n$ ranging from one to hundreds of thousands),
it is more convenient to treat $n$ as a continuous random variable
and estimate its empirical probability density using logarithmic binning 
(for visualization purposes only)
\cite{Corral_Deluca};
so $f(n)$ denotes in fact a probability density, as well as its empirical estimation.

The number of different types present in a dataset (the types with $n \ge 1$)
is what we call the vocabulary of the dataset, denoted by $V$
(this is bounded by $2^{12}=4096$).
The sum of all the frequencies of all types yields
the total number of tokens, which corresponds, by construction, 
to the dataset length $L$ measured in terms of the elementary time unit 
(number of beats, by default).
In a formula, $\sum_{i=1}^V n_i =L$, where $i$ labels the types.

\section{Power-law, Double Power-law and Lognormal Fits}

\subsection{Probability Densities and Rescaling}

\begin{figure}[t]
\includegraphics[width=.55\columnwidth]{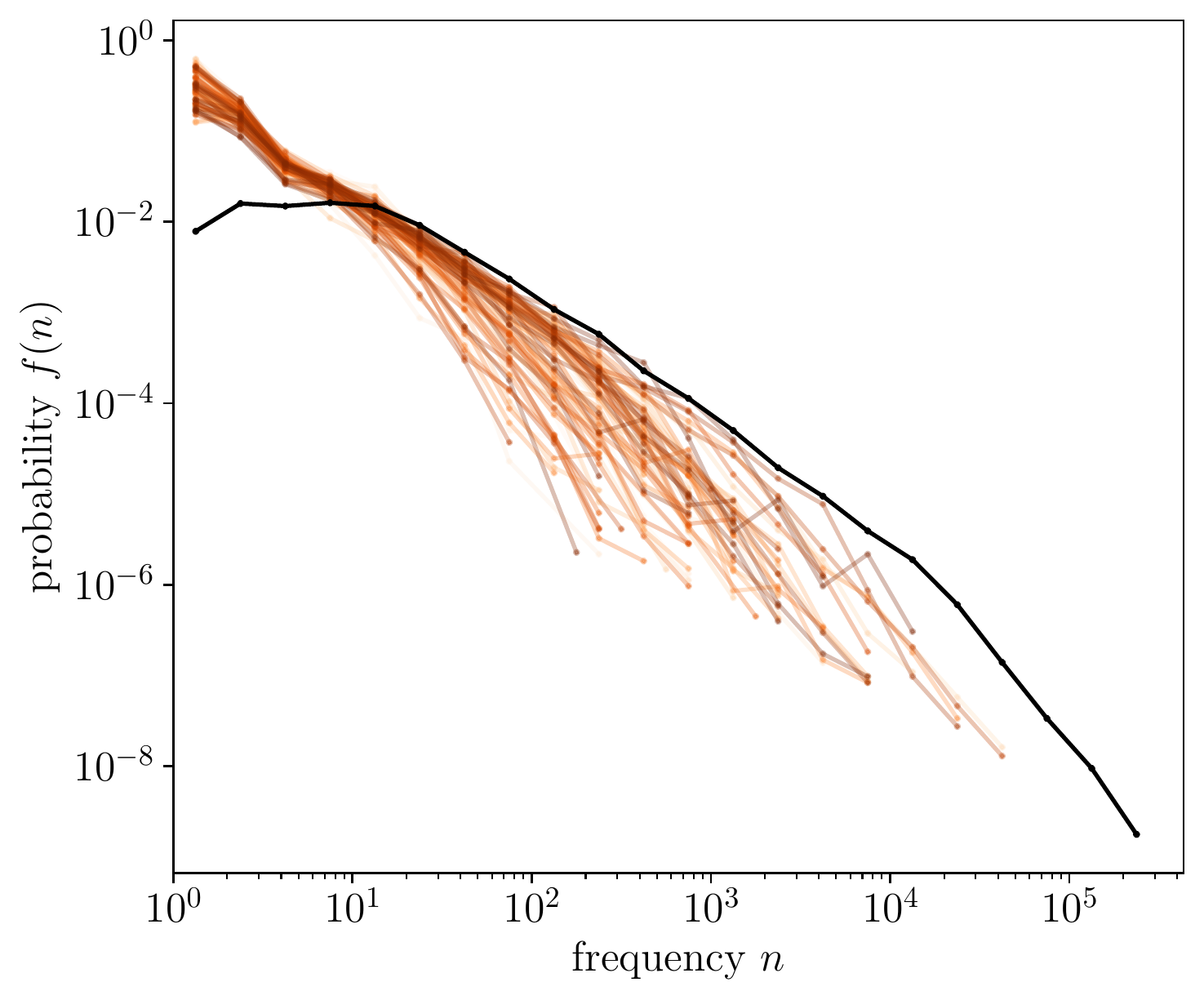}
\includegraphics[width=.55\columnwidth]{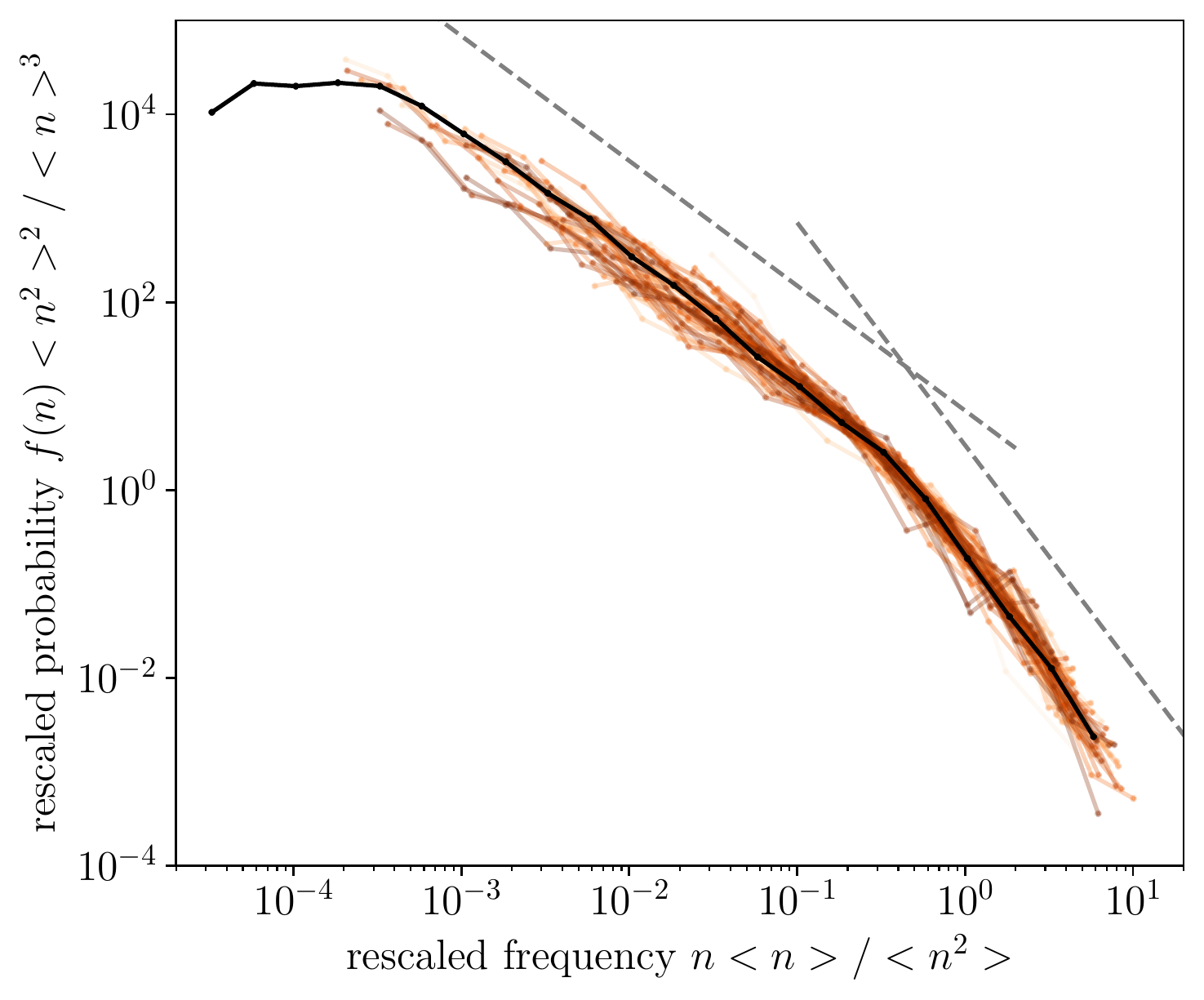}
\caption{
(a) Empirical probability densities of codeword frequency for the 76 composers, 
individually (orange-brown lines), and with all of them aggregated (black line).
(b) Same distributions under rescaling of the axes,
{where a roughly similar shape emerges for every composer and the global corpus.
Dashed straight lines are power laws with exponents $1.33$ and $2.4$,
derived from the fit of the global dataset. }
}
\label{fig_rosso}
\end{figure}

As a summary of the empirical probability densities of type frequency,
Fig. \ref{fig_rosso}(a) shows all of them
(77 in total, one for each composer plus the global one in which all composers are aggregated).
All distributions present many types that only occur once ($n=1$, the so-called hapax legomena in linguistics \cite{Baayen}),
as well as types with very high frequencies ($n>10^5$ in the global dataset),
with a rather smooth decaying curve linking both extremes.

When rescaled, a roughly similar shape is unveiled,
as seen in Fig. \ref{fig_rosso}(b).
Rescaling is done in the following way:
$n \rightarrow n \langle n \rangle / \langle n^2 \rangle$
and
$f(n) \rightarrow f(n) \langle n^2 \rangle^2/\langle n \rangle^3$,
with $\langle n\rangle$ and $\langle n^2\rangle$
denoting the first (mean) and second empirical moments of the distribution. 
{The reason behind such rescaling is the assumption of a hypothetical scaling form for 
$f(n)$, 
$$
f_\text{sca}(n)=\frac 1  a \left(\frac a \theta \right)^{\beta_1} G\left(\frac n \theta \right),
$$
defined for $n\ge a$, 
with 
$a$ a (fixed) lower cutoff, 
$\theta$ a scale parameter, and $G$ a scaling function behaving as a decreasing power law
with exponent $\beta_1$ for small arguments
($1<\beta_1<2$)
and decaying fast enough for large arguments
(in order that $\langle n\rangle $ and $\langle n^2 \rangle $ exist).
Then, 
$\langle n \rangle \propto a (\theta/a)^{2-\beta_1}$
and
$\langle n^2 \rangle \propto a^2 (\theta/a)^{3-\beta_1}$, 
and isolating, 
$\theta \propto \langle n^2 \rangle / \langle n \rangle$
and
$\theta^{\beta_1} \propto a^{\beta_1-1} \langle n^2 \rangle ^2/ \langle n \rangle^3$,
which justifies the rescaling
and allows its verification without knowledge of the values
of $\theta$ and $\beta_1$,}
see Ref. \cite{Corral_csf}.
When a literary piece is broken into different parts, 
this rescaling is equivalent to 
$n \rightarrow n /L$
and
$f(n) \rightarrow f(n) L V$ (see Ref. \cite{Corral_Font_Clos_PRE17});
however, this is not the case here, and the latter scaling form
(in contrast to the former one)
does not work well (not shown). 
Visual inspection of the rescaled plot based
on the ratio of moments (Fig. \ref{fig_rosso}(b))
suggests that the lognormal and the double power law
seem appropriate candidate distributions to fit the data.

In this work,
we consider three different fitting distributions, 
all of them continuous; 
thus, the frequency $n$ is assumed to be a continuous random variable.
The explanation of the three fitting distributions follows,
with special emphasis in the double power law.

\subsection{Simple Power-Law Distribution}

The simple power-law distribution, referred to as
untruncated power law or simply as power-law (pl) distribution, 
has a probability density
$$
f_\text{pl}(n)=     \frac{\beta-1}{a}
\left(\frac a n \right)^{\beta} \mbox{ for } n\ge a,
$$ 
and zero otherwise. The exponent $\beta$ fulfils $\beta > 1$
and $a$ is the lower cut-off, fulfilling $a>0$.
Considering $a$ as fixed, there is only one free parameter, which is $\beta$.
In the particular case in which 
$\beta$ is in the range $1.8 \le \beta \le 2.2$
we will talk about the fulfillment of Zipf's law.

\subsection{Double Power-Law Distribution}

The second fitting distribution is the one we call the double power-law (dpl) distribution \cite{Corral_Gonzalez},
whose probability density is given by
$$
f_\text{dpl}(n)=(1-q) \frac{\beta_1-1}{\theta} \frac 1 {c^{1-\beta_1}-1} \left(\frac \theta n \right)^{\beta_1} \mbox{ for } a\le n \le \theta,
$$
$$
f_\text{dpl}(n)=     q \frac{\beta_2-1}{\theta}                                     \left(\frac \theta n \right)^{\beta_2} \mbox{ for } n\ge  \theta,
$$ 
and zero for $n<a$.
Naturally, the lower cut-off $a$ may take a different value than for the simple power law, 
although at this point we use the same symbol for simplicity.
The two exponents $\beta_1$ and $\beta_2$ fulfill
$-\infty < \beta_1 < \infty$ with $\beta_1\ne 1$ and $\beta_2> 1$; 
$\theta$ is a scale parameter fulfilling $\theta\ge a$; 
and
the lower cut-off $a$ fulfills
$a \ge 0$ if $\beta_1 < 1$
and
$a >0$     if $\beta_1 > 1$.
The auxiliary parameter $c$ is defined as
$c=a/\theta$
and the parameter $q$ is not free either but ensures continuity 
at $n=\theta$
between the two regimes,
leading to
$$
q=\frac{\beta_1-1} {(\beta_2-1) c ^{1-\beta_1} - (\beta_2-\beta_1)},
$$
which gives $0 < q \le 1$.
As the expressions that multiply $1-q$ and $q$ in $f_\text{dpl}(n)$ are normalized
in their respective ranges, 
$q$ turns out to be the fraction of probability 
contained in the range $n\ge \theta$. 
%
%
If $a$ is fixed, the free parameters are $\beta_1$, $\beta_2$, 
and $\theta$.
The usual (untruncated) power-law distribution is recovered either in the limits 
$\theta=a$ (equivalent to $q=1$)
or
$\beta_1=\beta_2 > 1$.

%


Note that the double power-law distribution has two contributions:
on the left ($n\le \theta$) 
we have a truncated (from above) power-law distribution, with weight $1-q$;
on the right ($n\ge \theta$) 
we have an untruncated power law, with weight $q$.
We will take advantage of this fact to fit the double power law to the empirical data,
fitting, separately, a truncated power law in the range $a\le n\le b$
(where we have redefined $\theta$ as $b$), 
and fitting an untruncated power law in $n\ge a_2$, 
(redefining $\theta$ as $a_2$)
\cite{footnote_marc}.
In each fit, 
$a_2$ and $b$ are 
fixed and considered different, in general.
%

%
The method used for the fit is, in both cases, 
the one explained in Refs. \cite{Corral_Deluca,Corral_Gonzalez}
(see the Supplementary Information).
If both fits are accepted in some range 
(in the sense that they cannot be rejected, 
with $p-$value greater than $0.20$)
and their ranges overlap 
(in the sense that the upper cut-off $b$ of the truncated power law
is above the lower cut-off $a_2$ of the untruncated power law, 
and both power laws cross each other), 
the double power-law fit is not rejected
{(provided $a\le 32$, see below).}
The resulting value of $\theta$ is given by the value of $n$ at which both fits cross, 
which turns out to be
$$
\theta=\left[\frac q {1-q} \, \frac {\beta_2-1}{\beta_1-1}\,
\left(\left(\frac a b\right)^{1-\beta_1}-1\right)
\frac{a_2^{\beta_2-1}}{b^{\beta_1-1}} \right]
^{\frac 1{\beta_2-\beta_1}}.
$$
In the case in which the double power-law fit works well, we expect
$a_2\simeq \theta\simeq b$ (but with $a_2\le \theta\le b$).
The replacements 
$b\rightarrow \theta$ and $a_2\rightarrow \theta$
can lead to small changes in the fitted values of $\beta_1$ and $\beta_2$; nevertheless, the good visual performance of the fits
allows us to disregard such changes.

Although we could have fitted the power laws in the discrete case 
\cite{Moreno_Sanchez,Corral_Cancho},
we have considered the continuous case instead, in order to compare on equal footing 
with the (truncated) lognormal distribution defined below, which is continuous.
For high enough values of $n$, the distinction between continuous 
and discrete random variables becomes irrelevant,
but not for small values of $n$. 

The sudden change of exponent of the double power law 
at $n=\theta$ may seem ``unphysical'',
but the distribution works quite well for the number of data we are dealing with
(in the last section we discuss an extension of the double power law that avoids this ``unphysicality'').
The case of interest for us is when $0 <\beta_1 < \beta_2$;
specifically, when $\beta_2$ is between 1.8 and 2.2, 
the resulting power-law tail is in correspondence with Zipf's law; 
then, we will refer to this particular case of a double power-law distribution as ``double Zipf'' \cite{Ferrer2001a}.

\subsection{Truncated Lognormal Distribution}

The third fitting distribution that we deal with is the (lower) truncated lognormal (ln),
whose probability density is
$$
f_\text{ln}(n)=
{\sqrt{\frac 2\pi}}
\left[
 \mbox{erfc}\left(\frac{\ln a -\mu}{\sqrt{2} \sigma}\right)
\right]^{-1}
\frac 1{ \sigma n}
\exp\left(-\frac{(\ln n-\mu)^2}{2\sigma^2}\right)
\mbox{ for } n\ge a
$$
and zero otherwise,
with $a\ge 0$ and $\mu$ and 
$\sigma$ the two free parameters
(being the mean and the standard deviation of the associated
untruncated normal distribution);
$\mbox{erfc}(y)=\frac 2 {\sqrt{\pi}} \int_y^\infty e^{-x^2} dx$ is the complementary error function.
%
The fitting procedure \cite{Corral_Gonzalez} 
proceeds in exactly the same way as for the untruncated power law
(with the only difference that the lognormal involves two free parameters, $\mu$ and $\sigma$,
when $a$ is considered fixed).

\subsection{Fitting Method and Model Selection}

The random variable to fit is the absolute frequency $n$ of the codewords
(types); this choice is not obvious in Zipfian systems, 
see Ref. \cite{Corral_Cancho}. 
The fitting method is the one in Refs. \cite{Corral_Deluca,Corral_Gonzalez},
consisting in maximum-likelihood estimation
and the Kolmogorov-Smirnov goodness-of-fit test
for different values of the lower cut-off $a$
(and the upper cutoff $b$ for the first regime of the double power law).
The selected value of $a$ (and $b$) is 
the one that 
yields the largest number of types
in the fitting range 
{(i.e., the one comprising more realizations of the random variable)}
among all the fits that lead to a
$p-$value larger than 0.20.
If the resulting value of $a$ is larger than $10^{3/2}\simeq 32$
(corresponding to more than 1.5 orders of magnitude 
from $n=1$ to $a$),
the fit is rejected, otherwise it is ``accepted''
and considered a ``good fit''.

For the comparison of the fits provided by the different distributions,
we have to deal with  different subsets of the data 
(as the values of the lower cut-off $a$ will be different in each case, in general).
We take the simple criterion of selecting the fitting distribution that yields the smaller value of $a$, 
i.e., the model that explains the larger portion of the data.
The result is what we refer to as the ``best fit''.
This is explained in detail in the Supplementary Information.

\section{Results}

\subsection{Simple Power-Law and Double Power-Law Fits}

\begin{figure}[t]
\includegraphics[width=.55\columnwidth]{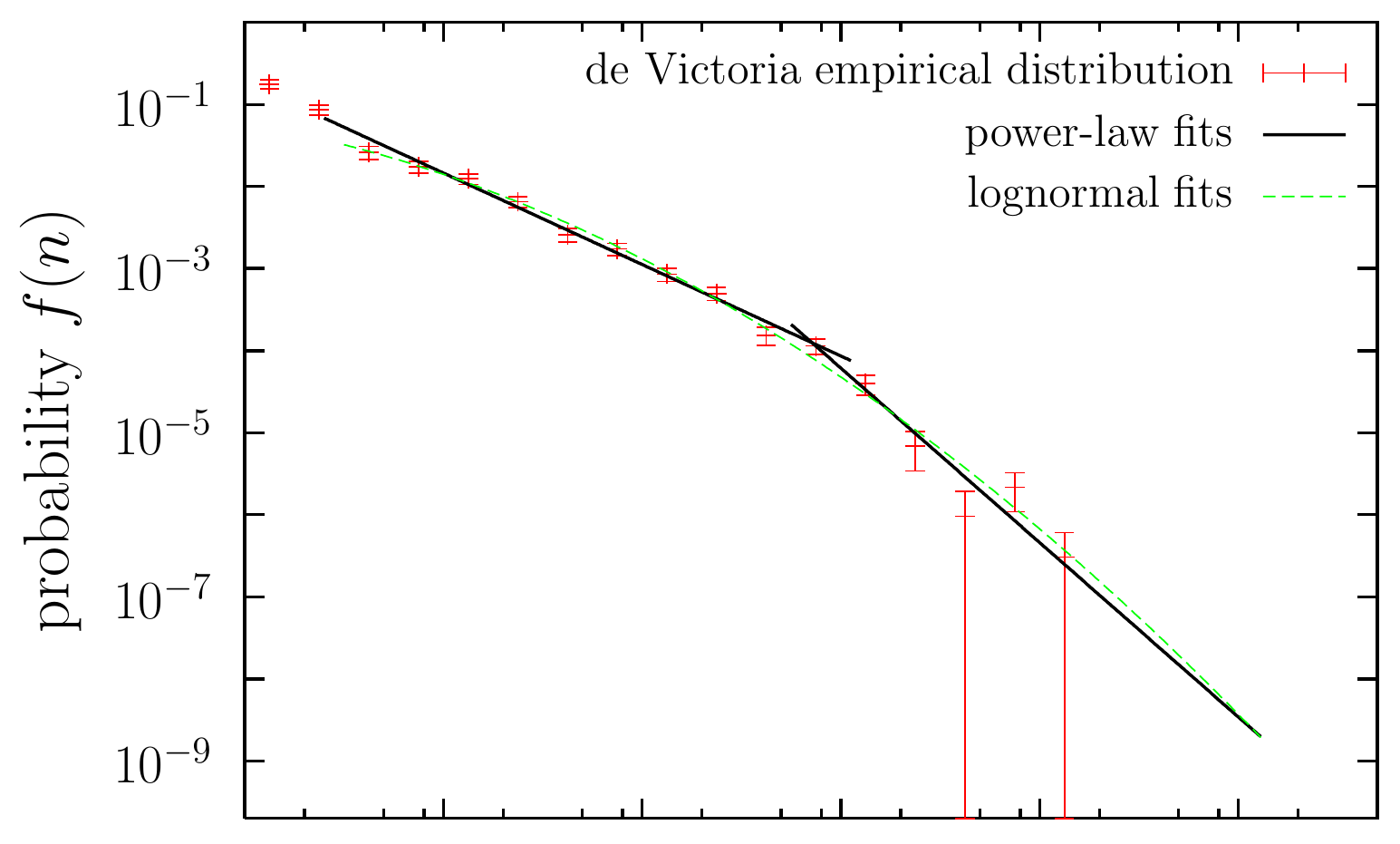}
\includegraphics[width=.55\columnwidth]{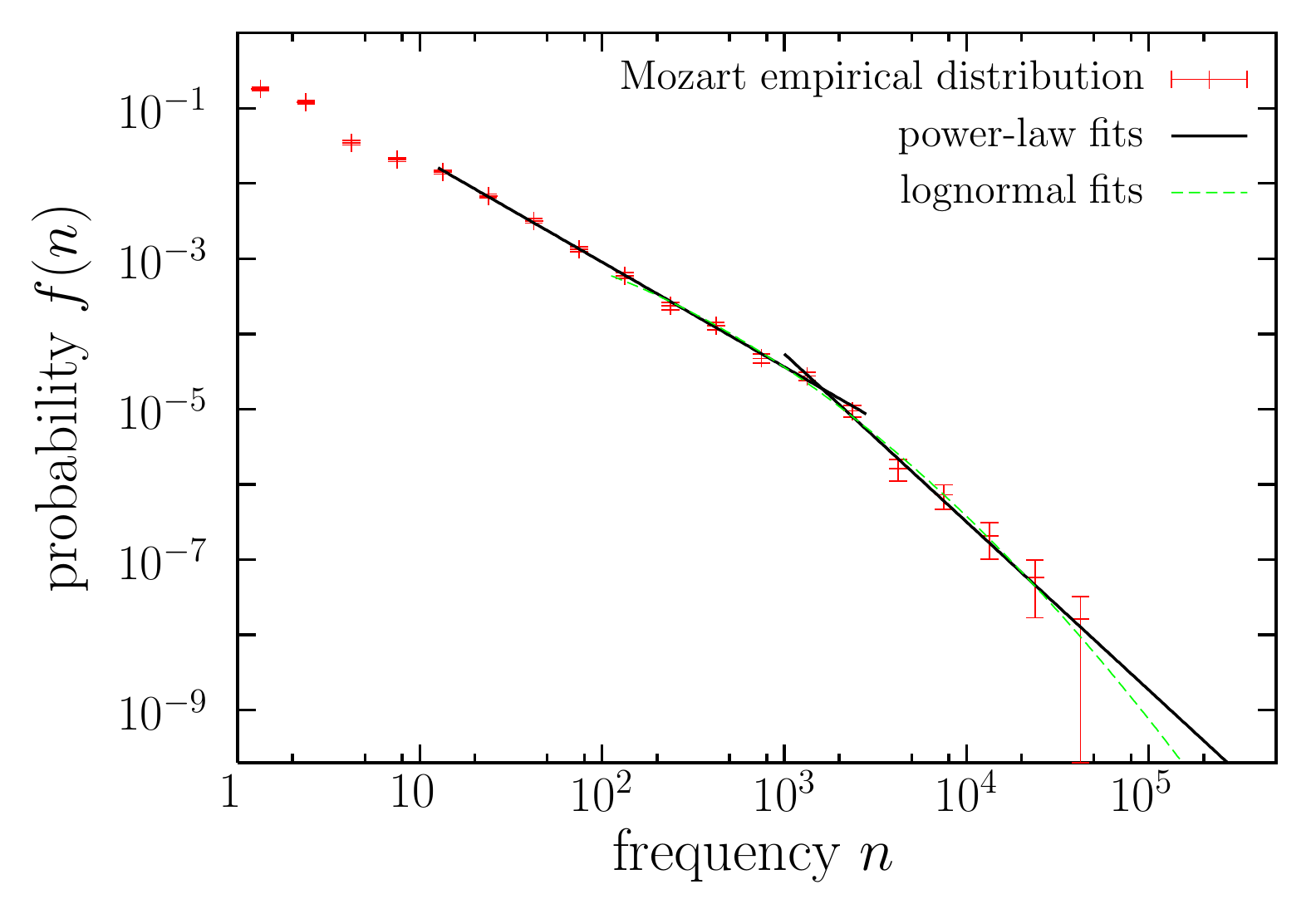}
\caption{
Empirical probability density of codeword frequency 
for (a) de Victoria and (b) Mozart, 
together with
double power-law and lognormal fits.
Among all the composers, 
de Victoria and Mozart yield the largest logarithmic span of the fitting range, $n_{max}/a$, 
for the double power law.
Note that de Victoria also yields the lognormal fit with the largest 
$n_{max}/a$
(comparable to the value for the double power law).
}
\label{fig_victoria}
\end{figure}

We start by comparing the results of fitting a simple power law
and a double power law.
We find that for the majority of the composers (48 of them, 63~\%) 
the double power law provides a good fit, 
which is obviously preferable to the simple power law
(due to the fact that
the double power law covers
a larger range of data than the simple power law,
as the latter is a part of the former).
In fact, when the double power law fits the data, 
the simple power law is rejected, as this only fits the tail, 
which is a too small fraction of the data (with $a_\text{pl}\gg 32$).
A couple of particularly good double power-law fits are shown as an illustration in Fig. \ref{fig_victoria},
where empirical probability densities and their fits are plotted together;
{there, it is also clear how the simple power law fits a rather small part of the data
($n>a_\text{pl}\simeq 10^3$)}.

\begin{figure}[ht] 
\includegraphics[width=.50\columnwidth]{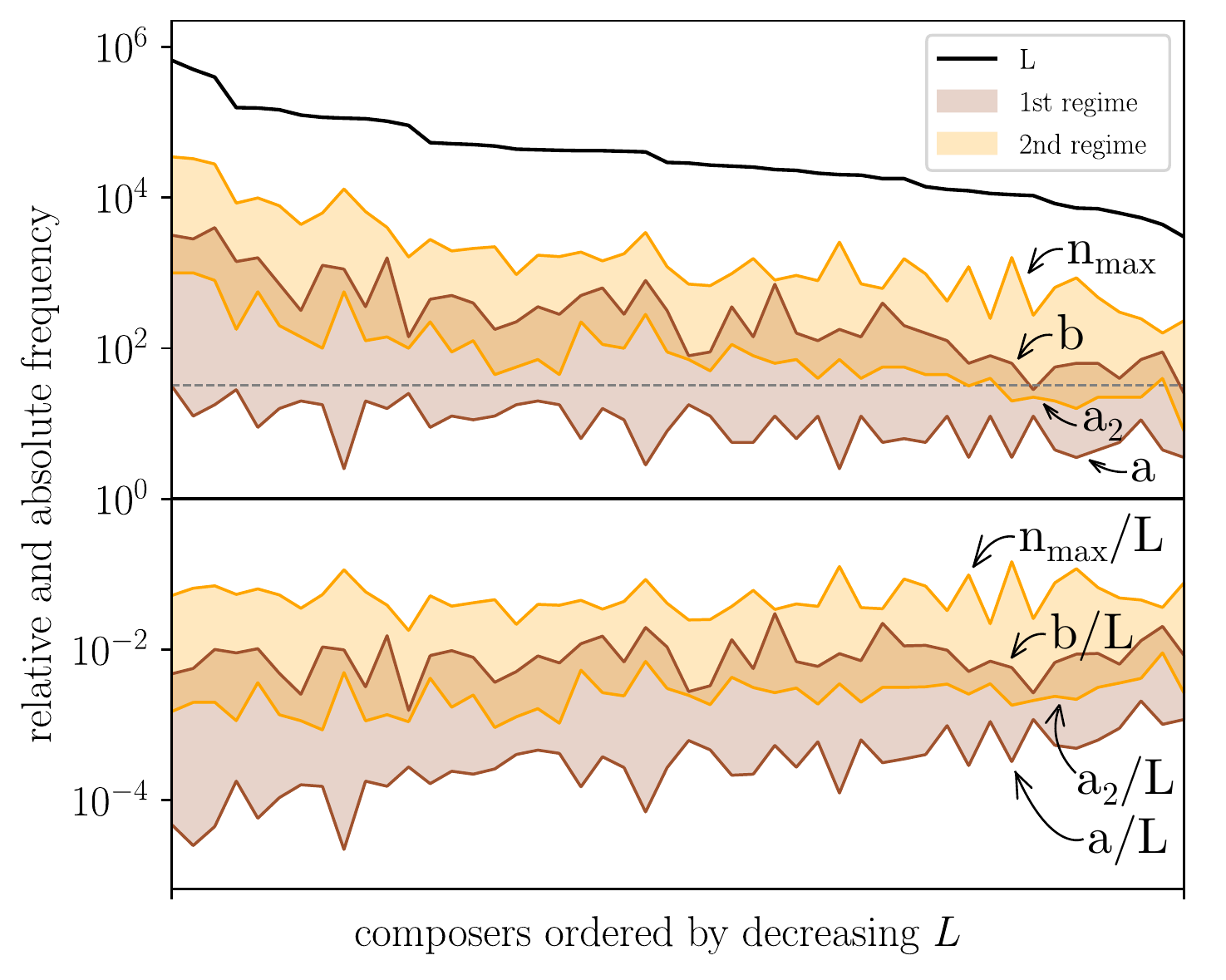}
\includegraphics[width=.47\columnwidth]{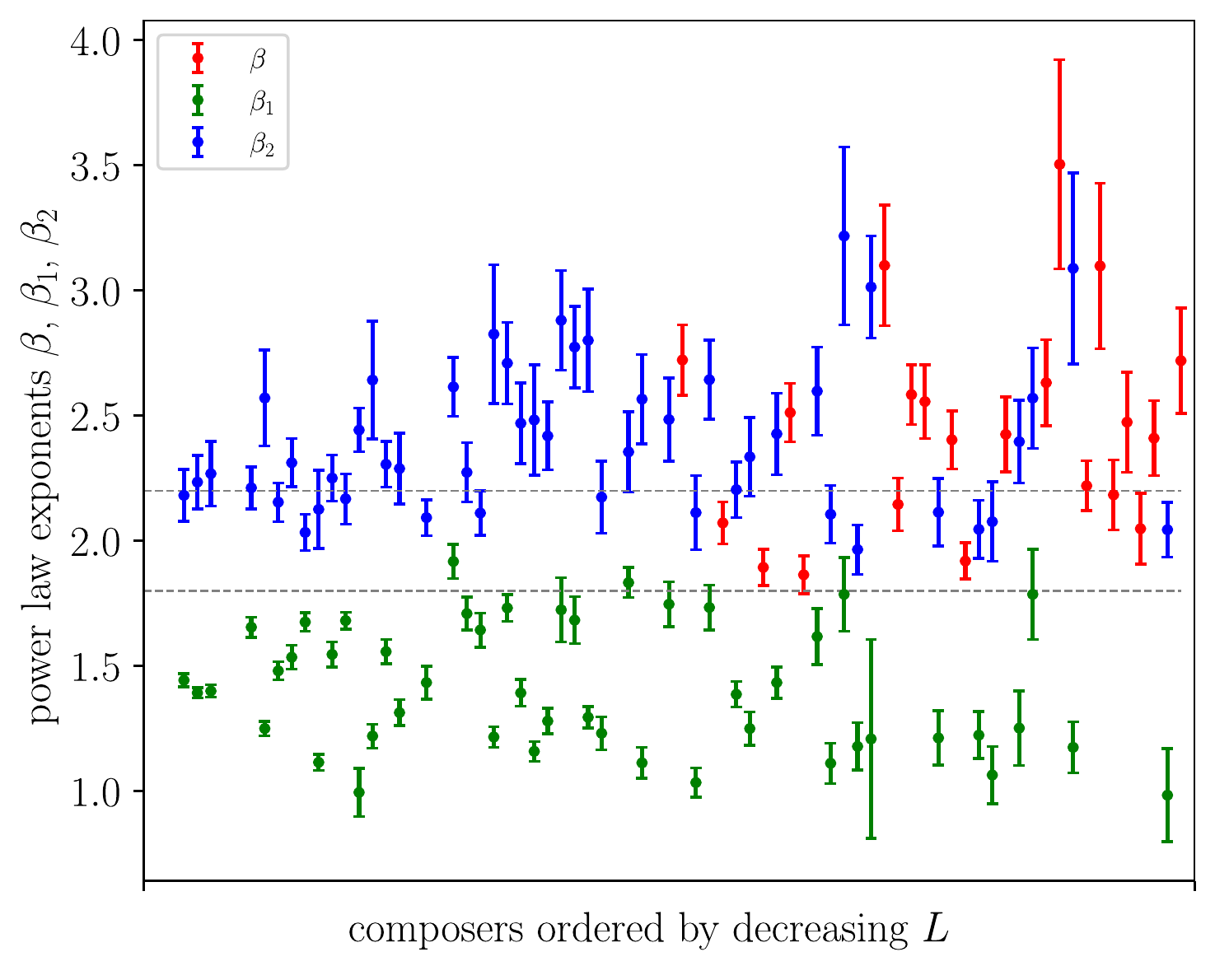}
\caption{
{Results 
of the double power-law fit
for each composer for which this is not rejected.}
{(a) Fitting ranges sorted by decreasing $L$.
}
Relative fitting ranges, 
in terms of cut-off frequencies divided by $L$, are also shown below.
The dashed line marks the $10^{1.5}\simeq 32$ limit.
The middle region not in the legend corresponds to the overlap 
between both regimes.
(b) Double power-law exponents $\beta_1$ and $\beta_2$.
The value of the exponent $\beta$ for the simple power law 
{(when this is not rejected)}
is also included.
Horizontal dashed lines delimit the Zipf's range for the
exponents $\beta_2$ and $\beta$. 
}
\label{fig_dpl}
\end{figure}

Figure \ref{fig_dpl} shows,
for each composer
for which the double power-law fit is not rejected,
the corresponding fitting ranges 
and exponents.
We see that when the cut-offs are expressed in terms of the relative frequency,
these are much more stable between different composers than when expressed in absolute frequencies 
(except for the relative minimum cut-off, $a_\text{dpl}/L$).
As a rough summary of the figure, 
the relative scale parameter of the dpl is $\theta/L \approx 0.005$,
and the maximum relative frequency is $n_{max}/L \approx 0.05$
(both with considerable dispersion). 
Curiously, $0.05$ is also, approximately, 
the relative frequency of the most common word in English, which is ``the'' \cite{Ngramviewer}. 

We also see in the figure that
the exponent 
$\beta_1$ ranges between 1 and 2, 
for most of the 
composers. 
{This power-law regime coincides qualitatively with what has been found in linguistics, 
using large corpora (where the exponent $\beta_1$ seems to be between 
1.4 \cite{Montemurro01,Corral_brevity}
and 
1.6 \cite{Ferrer2001a,Petersen_scirep,Gerlach_Altmann}),
but we are not aware that it was reported before in music.
}
In addition, we observe that 
$\beta_2$ ranges mostly between 2 and 3, 
(remember that in order to consider that we have a Zipfian tail, $\beta_2$ should be between 1.8 and 2.2, roughly).
As the composers are ranked by decreasing $L$,
we also observe that 
the smaller $L$, 
the larger the dispersion in the values of $\beta_2$ and $\beta$.

However, there are a number of cases (28, 37~\%) in which the double power-law fit is not appropriate, 
and this can be due to two main reasons: 
either the two power-law regimes do not overlap (6 composers)
and thus the fit is rejected,
or the first power-law regime 
is meaningless and thus the fit is also rejected.
The latter can arise from two subcases: 
from a too short fitting range (2 composers),
or from the fact that both exponents are nearly the same, i.e., $\beta_1\simeq\beta_2$, 
and then 
the existence of two power-law regimes cannot be established
(20 composers).
%

%
%

\begin{figure}[ht]
\includegraphics[width=.55\columnwidth]{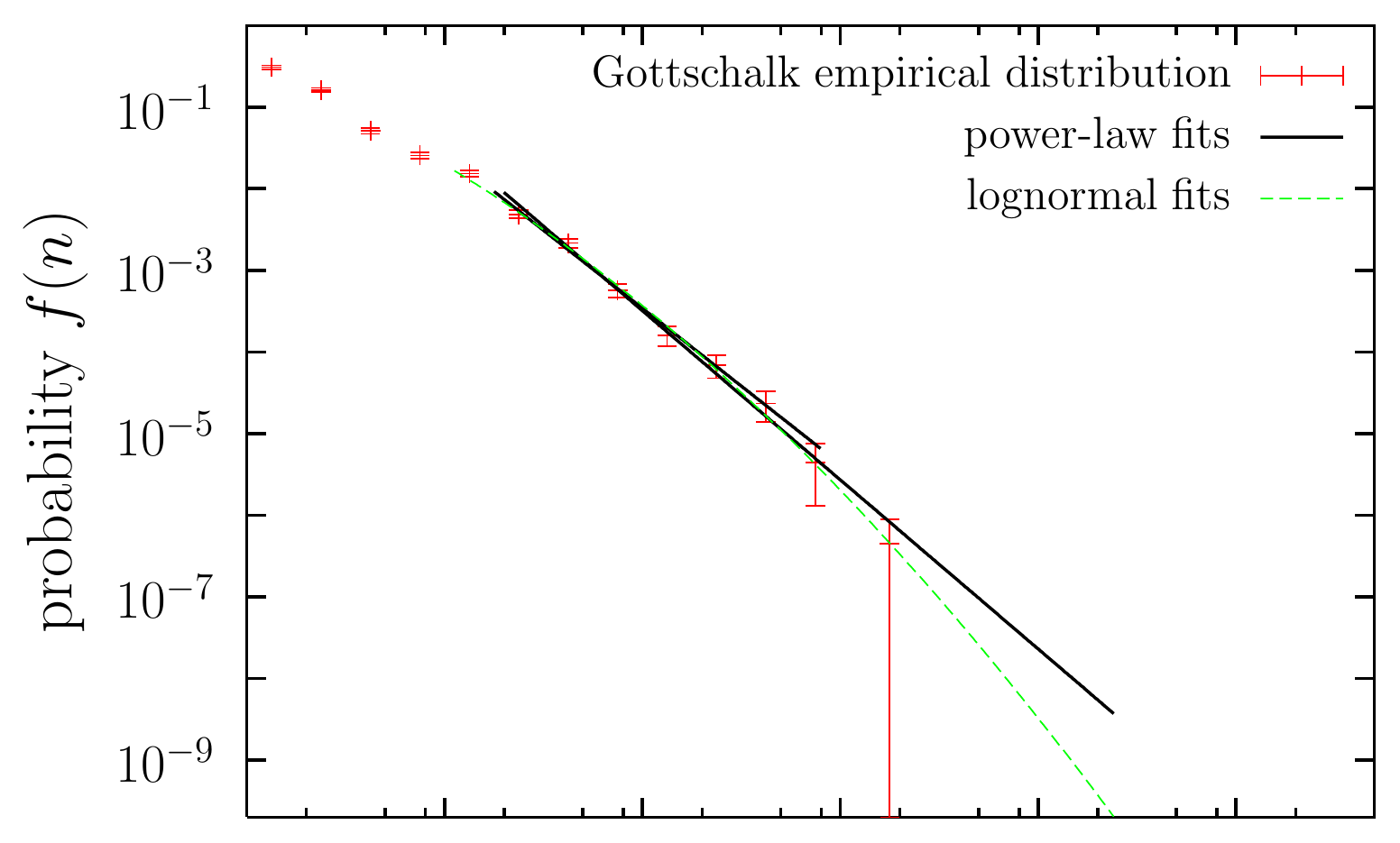}
\includegraphics[width=.55\columnwidth]{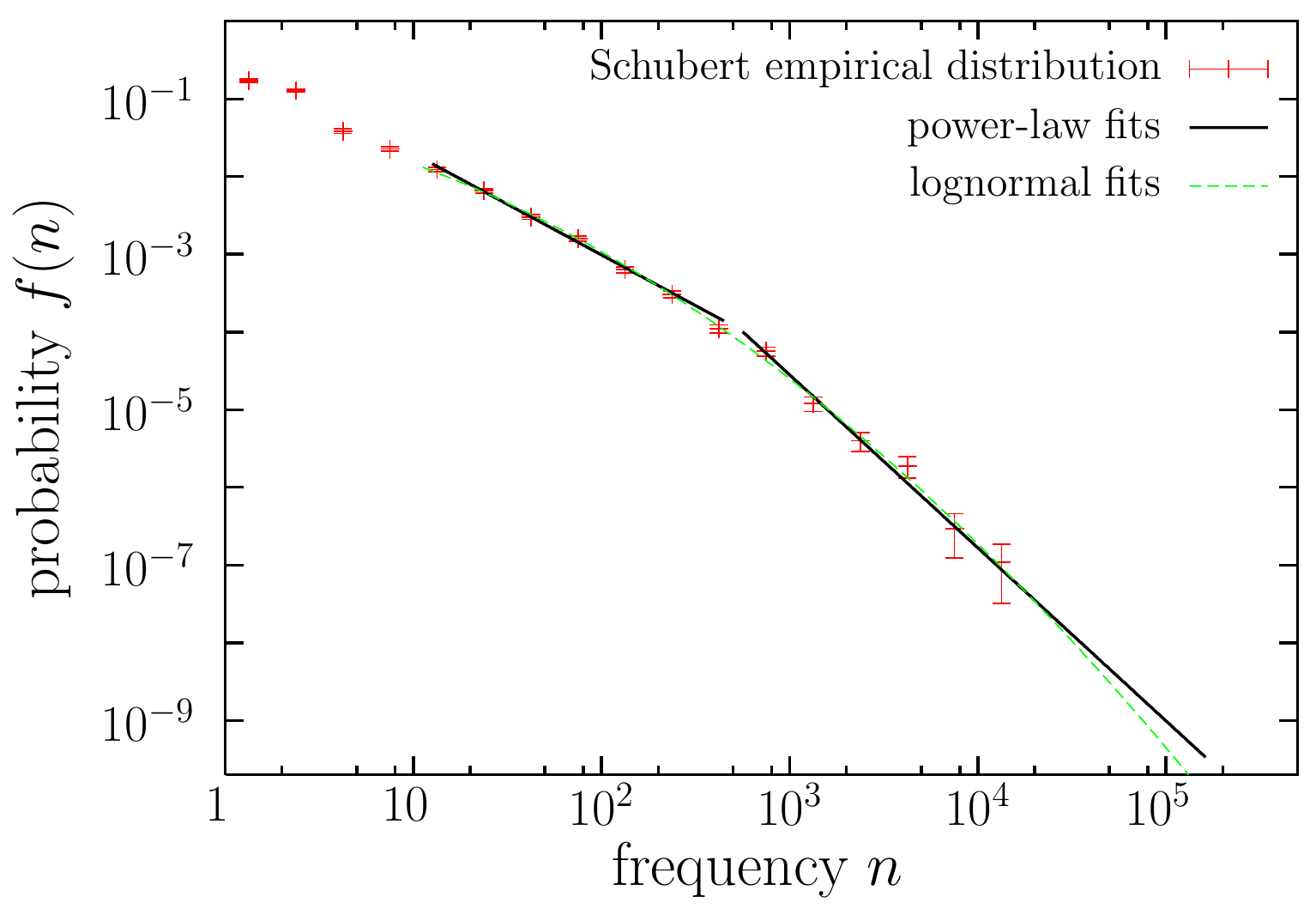}
\caption{
Empirical probability density of codeword frequency 
for (a) Gottschalk and (b) Schubert, 
together with their fits.
Gottschalk is the composer with the largest logarithmic fitting range for the simple power law, 
nevertheless, the (logarithmic) fitting range for the lognormal is larger.
Schubert has the largest number of types in the lognormal fitting range (1102).
Failed double-power law fits are also shown.
}
\label{fig_schubert}
\end{figure}

%
Table \ref{tableone}
displays the results 
for these two subcases, clearly showing the failure of the double power law
($b/a_\text{dpl}$ small or $\beta_1\simeq \beta_2$; the table makes these statements quantitative).
The table also shows that,
when the first power-law regime (the truncated one)
is meaningless,
the simple power law provides a good fit 
for all composers but one
(21 composers; 
Bruckner, with $a_\text{pl}=35$, is excluded).
In most of the cases,
the fitted power-law exponent ranges from $\beta\simeq 1.9$ to 2.7
(also displayed at Fig. \ref{fig_dpl}).
Figure \ref{fig_schubert}(a) shows 
one case of the failure of the double power-law fit
and the validity of the simple power law.

\begin{table}[h]
\begin{center}
\caption{\label{tableone}
Results of the double power-law fit when this fit is rejected, 
either because the truncated power-law regime has a too short range
or because there are not two different power-law regimes
(case of non-overlapping fitting ranges is not included, 
except for the global dataset) \cite{footnote_marc}.
The simple power law fit is accepted (with $\beta=\beta_2$) for all composers but one (Bruckner, for which $a_\text{pl}=a_2=35$).
The number of orders of magnitude of this fit is given by $\ell=\log_{10}(n_\text{max}/a_2)$.
The number of types included in the fit is $v_2$.
The uncertainties in $\beta_1$ and $\beta_2$ are given by the standard deviation of the maximum likelihood estimation.
The rest of variables are explained in the main text.
The Zipfian subcases ($1.8 \le \beta \le 2.2$) are marked in bold.
}
\smallskip
\tiny
\setlength{\tabcolsep}{10pt} 
\begin{tabular}{l rrr c  rrcrc}
Composer	&	$V$	&	$a$	&	$b$	&	$\beta_1$		&	$a_2$	&	$n_\text{max}$	&	$\ell$	&	$v_2$	&	$\beta_2$			\\
\hline
{\bf Monteverdi} 	&	247	&	8	&	319	&	1.895	$\pm$	0.126	&	10	&	319	&	1.50	&	68	&	{\bf 2.048	$\pm$	0.142}	\\
{\bf Scheidt} 	&	233	&	11	&	475	&	2.105	$\pm$	0.155	&	11	&	475	&	1.63	&	76	&	{\bf 2.184	$\pm$	0.139}	\\
Albrechtsberger	&	511	&	16	&	282	&	2.634	$\pm$	0.205	&	16	&	282	&	1.25	&	84	&	2.632	$\pm$	0.172	\\
{\bf Paganini}	&	713	&	10	&	351	&	1.638	$\pm$	0.097	&	8	&	351	&	1.65	&	166	&	{\bf 1.919	$\pm$	0.072}	\\
Cramer	&	657	&	11	&	56	&	1.786	$\pm$	0.249	&	20	&	218	&	1.04	&	50	&	3.098	$\pm$	0.330	\\
{\bf Berlioz}	&	805	&	11	&	398	&	1.844	$\pm$	0.097	&	18	&	602	&	1.53	&	109	&	{\bf 2.146	$\pm$	0.104}	\\
{\bf Gottschalk}	&	1035	&	18	&	794	&	1.908	$\pm$	0.086	&	20	&	2410	&	2.08	&	174	&	{\bf 2.072	$\pm$	0.084}	\\
{\bf Bizet}	&	782	&	11	&	751	&	1.709	$\pm$	0.080	&	13	&	751	&	1.78	&	155	&	{\bf 1.864	$\pm$	0.075}	\\
Bruckner	&	1602	&	40	&	891	&	2.338	$\pm$	0.157	&	35	&	895	&	1.40	&	132	&	2.436	$\pm$	0.119	\\
Guilmant	&	522	&	9	&	276	&	2.349	$\pm$	0.159	&	9	&	276	&	1.49	&	92	&	2.410	$\pm$	0.149	\\
Janáček	&	874	&	16	&	251	&	2.323	$\pm$	0.148	&	18	&	251	&	1.15	&	125	&	2.556	$\pm$	0.147	\\
Joplin	&	634	&	8	&	325	&	2.154	$\pm$	0.116	&	8	&	325	&	1.61	&	119	&	2.220	$\pm$	0.099	\\
{\bf Satie} 	&	912	&	11	&	781	&	1.728	$\pm$	0.073	&	13	&	781	&	1.79	&	180	&	{\bf 1.894	$\pm$	0.073}	\\
Karg-Elert	&	752	&	9	&	101	&	2.594	$\pm$	0.225	&	9	&	101	&	1.05	&	75	&	2.720	$\pm$	0.210	\\
Respighi	&	891	&	16	&	348	&	2.304	$\pm$	0.139	&	16	&	348	&	1.34	&	122	&	2.404	$\pm$	0.116	\\
Rajm\'aninov	&	2231	&	13	&	380	&	2.375	$\pm$	0.099	&	18	&	380	&	1.33	&	146	&	2.512	$\pm$	0.117	\\
Schoenberg	&	997	&	10	&	335	&	2.261	$\pm$	0.212	&	9	&	335	&	1.58	&	68	&	2.474	$\pm$	0.200	\\
Bart\'ok	&	1766	&	13	&	226	&	2.503	$\pm$	0.136	&	13	&	226	&	1.25	&	153	&	2.584	$\pm$	0.119	\\
M\'edtner	&	1233	&	13	&	74	&	2.825	$\pm$	0.295	&	16	&	74	&	0.67	&	49	&	3.504	$\pm$	0.418	\\
Gershwin	&	1827	&	16	&	50	&	2.398	$\pm$	0.291	&	22	&	154	&	0.84	&	81	&	3.100	$\pm$	0.241	\\
Prok\'ofiev	&	1171	&	13	&	291	&	2.345	$\pm$	0.164	&	13	&	291	&	1.36	&	93	&	2.426	$\pm$	0.149	\\
Stravinsky	&	2442	&	28	&	296	&	2.574	$\pm$	0.159	&	28	&	296	&	1.02	&	154	&	2.723	$\pm$	0.141	\\
\hline
All 76	&	4085	&	40	&	3980	&	1.334	$\pm$	0.017	&	10000	&	254000	&	1.40	&	118	&	2.369	$\pm$	0.120	\\
\end{tabular}
\par
\end{center}
\end{table}

\begin{figure}[ht]
\includegraphics[width=.55\columnwidth]{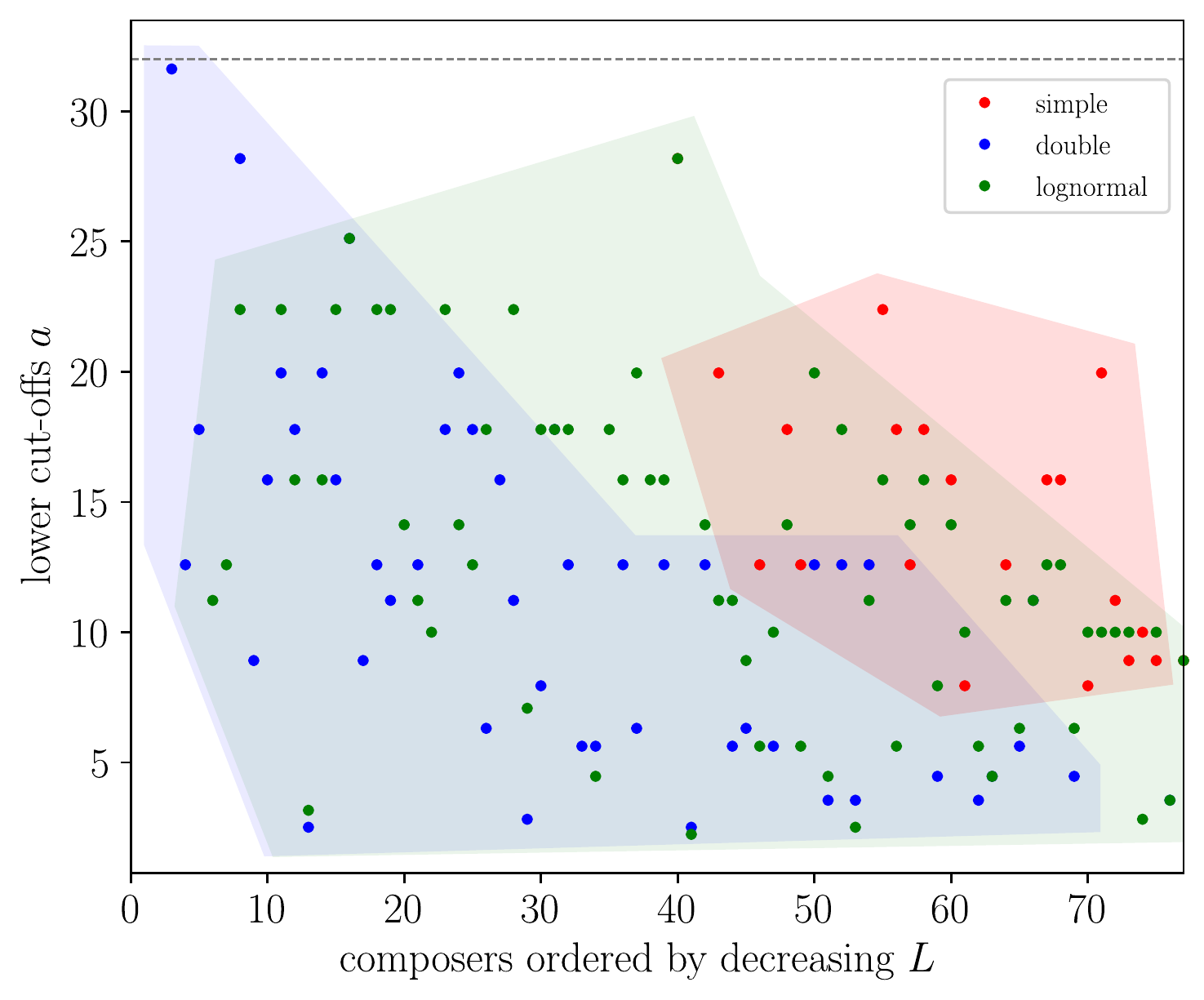}
\caption{
Results for each composer of the lower cut-off for each fit,
when this is not rejected.
{Shadowed regions group the results for each fit.}
A lower value of $a$ implies a larger fitting range;
the distribution with the lowest value of $a$
is the preferred best fit for each composer.
}
\label{fig_cutofss}
\end{figure}

Coming back to the case
when the double power law is rejected because no overlap between the two regimes exists 
(6 composers only, as Schubert in Fig.~\ref{fig_schubert}(b)),
the simple power law is rejected as well, as the value of $a_\text{pl}$ 
turns out to be too high ($a_\text{pl} \gg 32$) 
and the power-law fit only 
includes the tail of the empirical distribution. 
Nevertheless, we will see that
in 5 of these cases the lognormal provides a good fit
(see Fig. \ref{fig_schubert}(b));
the lognormal becomes the best fit then, 
as represented in Fig. \ref{fig_cutofss}.
The exception to this is given by Bach, 
who turns out to be the only composer for which
none of the three distributions is 
able to fit the data 
with $a\le 32$.
%
%

\begin{figure}[ht]
\includegraphics[width=.55\columnwidth]{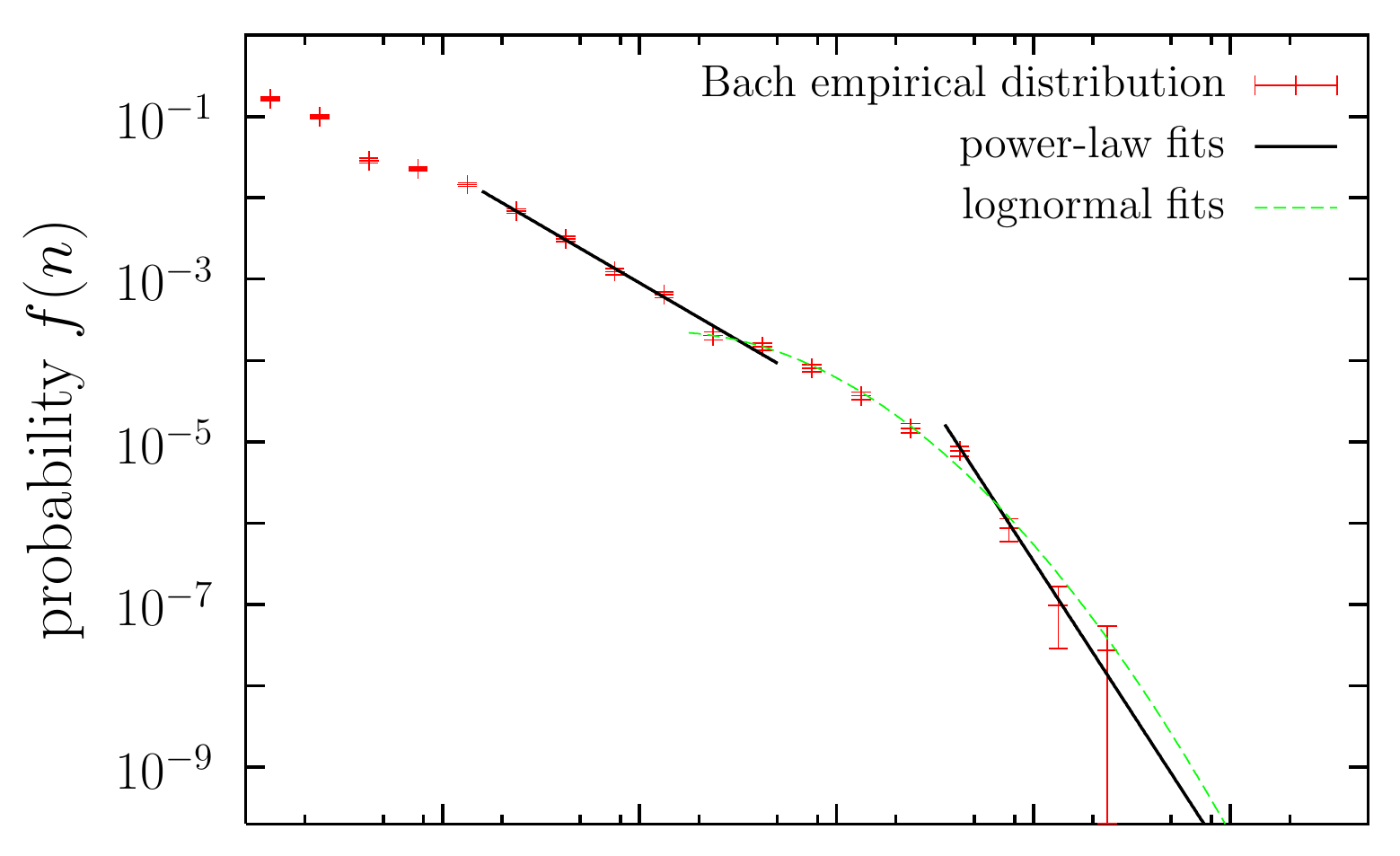}
\includegraphics[width=.55\columnwidth]{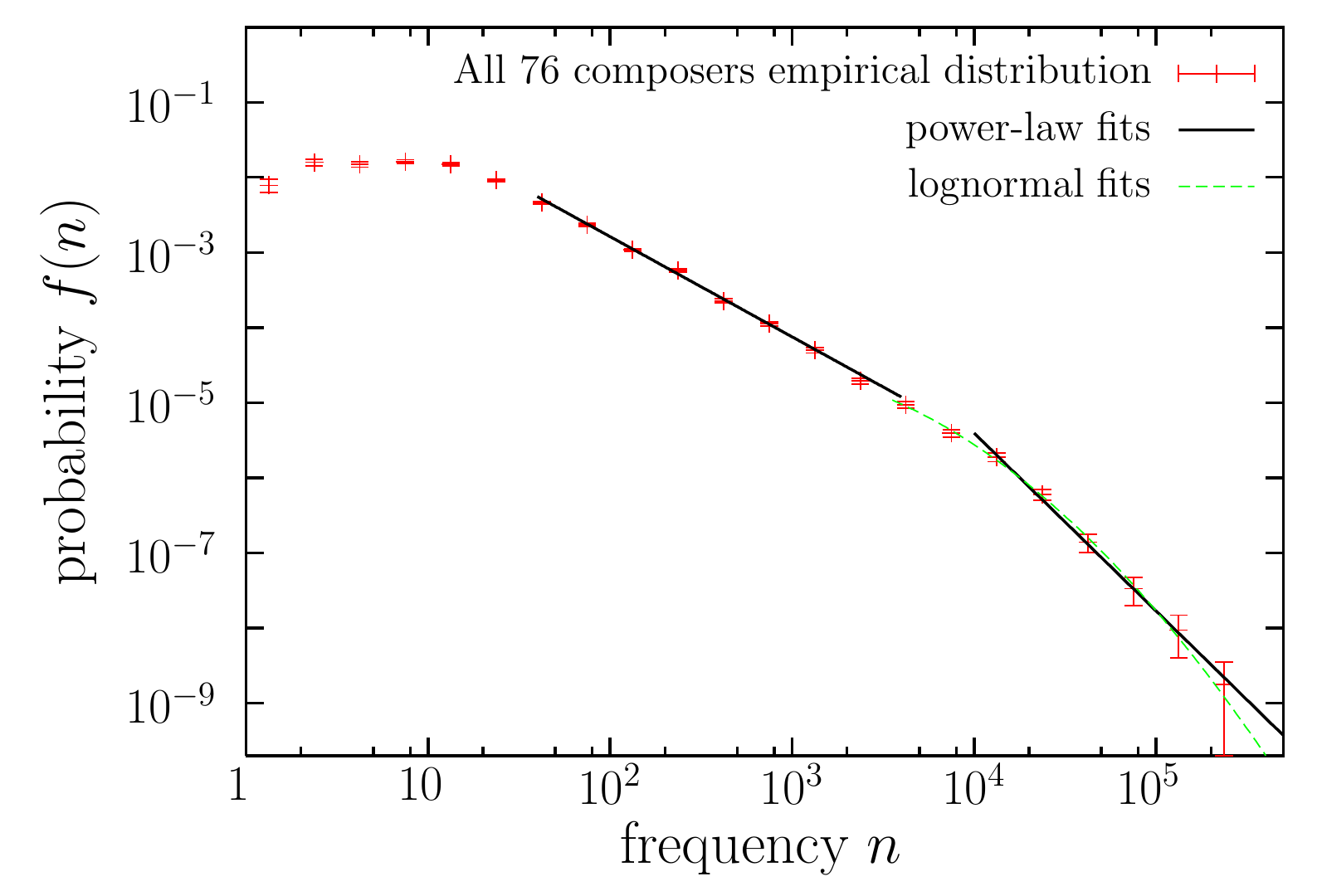}
\caption{
Empirical probability density of codeword frequency 
for (a) Bach, and
for (b) the global dataset with all the 76 composers, 
together with power-law fits and lognormal fit.
In both cases the double-power law fit is rejected, 
as the two fitted regimes do not overlap, 
and also the lognormal fit is unsatisfactory,
as it excludes a significant part of the types
(i.e., the value of $a$ turns out to be too large).
}
\label{fig_bach}
\end{figure}

We show Bach's empirical distribution of frequencies together with its different (failed) fits in Fig. \ref{fig_bach}(a);
there one can see that the two power-law regimes are far from overlapping,
and that the tail exponent is rather large ($\beta\simeq 3.7$)
and limited to a narrow range {in frequency}.
Note that Bach's distribution could be fitted by a power-law body followed 
(with overlap) by a lognormal tail, 
but such a distribution is not considered in this paper
(of course, other distributions with a supercritical bump could be considered as well \cite{Corral_garcia_moloney_font}).

As another example of a failed double power-law fit, due to the non-overlapping of the two regimes, 
we consider in Fig. \ref{fig_bach}(b) the global dataset of the 76 composers,
although it is remarkable that in this case the double power-law fit is 
not far from being ``accepted''
(the region of no overlap has a rather short range).
The two power law exponents would be 
$\beta_1=1.33$ and $\beta_2=2.4$
(this case has also been included in Table \ref{tableone}).

\subsection{Lognormal Fit}

Regarding the lognormal fit, it is rejected for only 9 composers (12~\%), due to a value of $a_\text{ln}$ too large (greater than 32).
{These are Couperin, Dandrieu, Bach, H\"andel, Haydn,
Mozart, Beethoven, Alkan, and Shostak\'ovich,
for which
the double-power law provides a good fit, except for Bach.}
The values of the cut-offs for the remaining 67 are shown in Fig. \ref{fig_cutofss}
{(this includes all cases that were not fitted by the double power law, except Bach).}
When comparing the double power law with the lognormal, 
we find that in many cases the two distributions provide roughly similar fits
(see for instance Fig. \ref{fig_victoria}(a);
{in Fig. \ref{fig_victoria}(b), despite some similarity,
the fitting ranges are clearly different).}
Out of the 40
composers that are fitted by both distributions,
for 25 of them the double power law provides the best fit
(in the sense that it fits a larger fraction of the data), 
and for 15 of them the situation is reversed (Fig. \ref{fig_cutofss}).

All composers fitted by the simple power law (21) are also fitted by the lognormal,
and
the fit provided by the lognormal is better for 14 of these 21 composers
(for the remaining 7 composers the simple power law is able to fit 
a larger range than the lognormal; this can be seen in Fig. \ref{fig_cutofss}). 
Figure \ref{fig_schubert}(a) is a good illustration of the case
when the lognormal outperforms the simple power law.
Figure \ref{fig_schubert}(b) shows another case, 
in which both the simple and the 
double power-law fail but not the lognormal. 

\section{Discussion and Conclusions}

Table \ref{tab:summary} summarizes our results. 
Recapitulating, 
the lognormal is the distribution that 
gives a good fit for more composers (67 out of 76, 88~\%),
followed by the double power law (48, 63~\%), 
and by the simple power law (21, 28~\%).
If one considers the simple power law as a special case of the double power law,
one would obtain a larger number for the latter distribution (69, 91~\%), 
thus accounting for roughly the same percentage of fits as the lognormal.
Note that
good simple power-law fits arise for composers
with low representativity in the corpus 
(low value of $L$ and smaller absolute frequencies);
so, we conjecture that increasing the number of pieces by these composers would unveil a range of smaller relative frequencies
which could be well fitted by another power-law regime
and thus a double power law would emerge in these cases
{(in other words, for the number of pieces in the corpus of these composers,
the value of $\theta$ would be too small to be detectable).
}

Among the double power-law fits, 
there are only 15 cases that can be considered good ``double Zipf'' fits
(with $1.8 \le \beta_2 \le 2.2$), 
and among the simple power-law fits, 
there are 7 with an exponent in the Zipf range.
So, out of 76 composers only 22 (29~\%) can be considered to follow Zipf's law the sense of a good fit 
(although a lognormal can provide a better fit for some of them).
{Considering best fits only, Zipf's law only arises for 9 composers
(12~\%).}

\begin{table}[h]
\caption{Summary of the fits for the 76 composers.
First row counts
distributions yielding the best fits (in terms of fitting more data);
Second row counts
the Zipf subcases among the best fits (exponents $\beta$ and $\beta_2$ between 1.8 and 2.2);
third counts
distributions yielding good fits (fit not rejected but not necessarily the best);
and fourth counts
the Zipf subcases among the good fits.
The statistics for the reason of rejection (bad fits) are also included
(in brackets).}
\label{tab:summary}
\setlength{\tabcolsep}{10pt}
\begin{tabular}{|lccc|} \hline
                 & Simple pl              & Double pl                                  & Lognormal            \\ \hline
Best fits        & 7                    & 33                                             & 35                   
\\
Best Zipf        & 1                    & 8                                             & --                   
\\
Good fits       & 21                   & 48                                             & 67                   \\
Good Zipf      & 7                    & 15                                             & --                    \\ \hline
Reason   & \multicolumn{1}{c}{$a_\text{pl}>32$} & 
\multicolumn{1}{c}{no overlap (6)} & 
\multicolumn{1}{c|}{$a_\text{ln} > 32$} \\
for rejection                 & \multicolumn{1}{c}{(55)} & 
\multicolumn{1}{c}{$b \approx a_\text{dpl}$ (2)}  & 
\multicolumn{1}{c|}{(9)} \\
                 & \multicolumn{1}{c}{} & 
\multicolumn{1}{c}{$\beta_1 \approx \beta_2$ (20)} & 
\multicolumn{1}{c|}{} \\ \hline
\end{tabular}
\end{table}

\begin{figure}[t]
\includegraphics[width=.55\columnwidth]{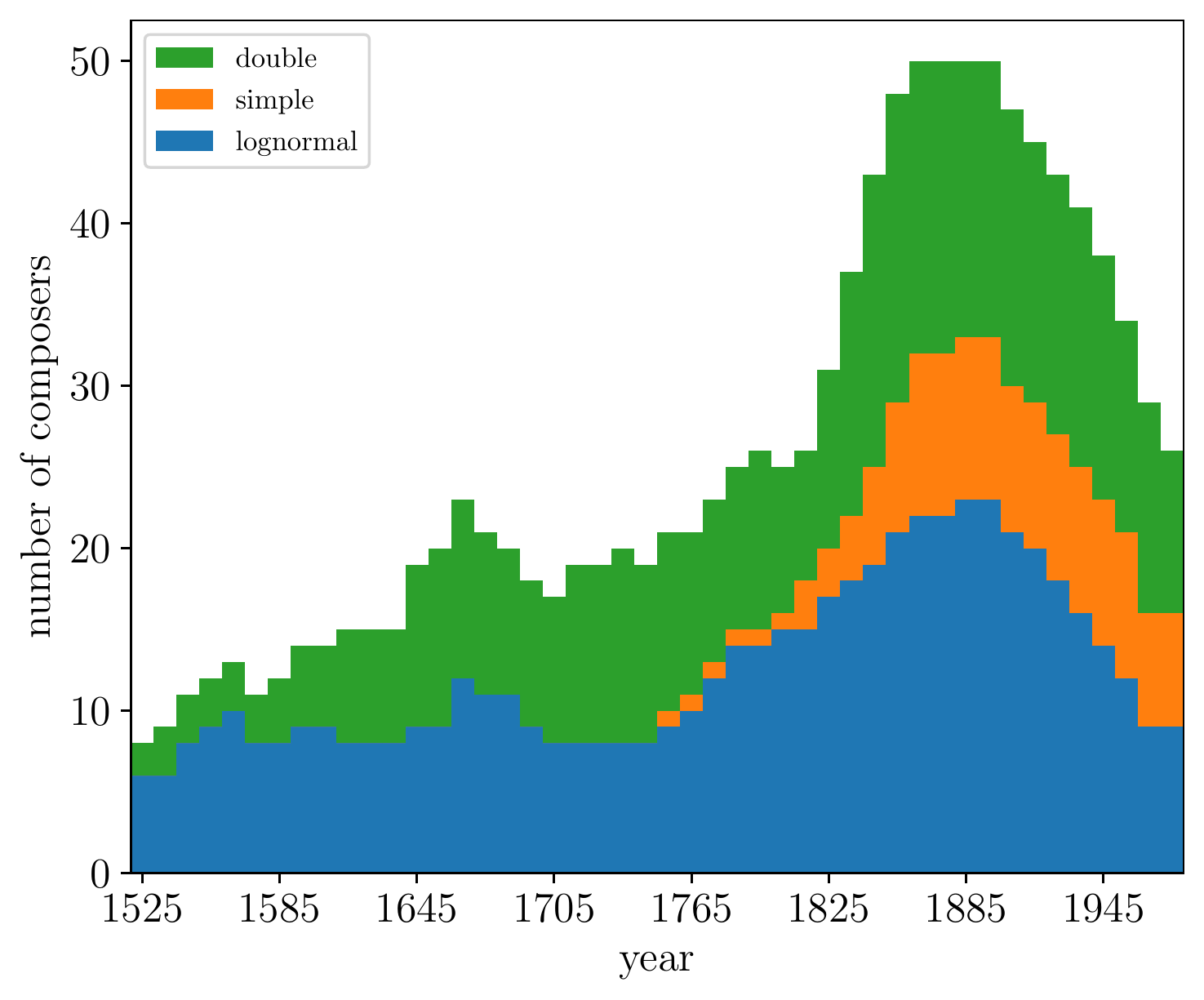}
\caption{
Number of composers best fitted by each fit, as
a function of year 
{(year is determined as the mean between birth plus 20 and death).}
Numbers have been smoothed: for each interval of 10 years we take a window of 150 years and count the number of fits in the window.
}
\label{fig_last}
\end{figure}

If we ask instead,
not which distributions fit well the data but
for the distribution that yields the best fit
{(in the sense of fitting more data, remember),
Table \ref{tab:summary} shows that}
the lognormal does it for 35 composers, 
the double power law for 33, 
and the simple power law for 7;
one case (Bach) is not fitted well by any of the three distributions.
Figure \ref{fig_last} quantifies these differences in chronological order.
{Except for the 15th century, the double power law
slightly dominates about the first half of the data,
but after Beethoven, the lognormal takes over.}
Interestingly, we observe that 
the simple power law only starts to appear as a best fit 
in the second part of the corpus, around the 18th century, 
and corresponds to composers with little representation in the corpus (low $L$).
Nevertheless, composers with low $L$ 
can also be found
in the first part of the corpus, but the simple power law does not provide the best fit in any of such cases.

For a small number of composers 
(six, as well as for the global aggregated dataset), 
the double power law is rejected because the two power-law regimes
do not overlap.
{This is an unfortunate situation, as the two power-law regimes exist}, but the double power law turns out to be too sharp 
in its transition from one power-law regime to the other at $n=\theta$.
{A smoother version of the double power law (sdpl) could be used instead,
$$
f_\text{sdpl}(n)= \frac \gamma {B \theta}
\left(\frac \theta n\right)^{\beta_1}
\left[\frac{1}{1+\left(n/\theta\right)^\gamma}\right]
^{\frac{\beta_2-\beta_1}\gamma},
\mbox{ for } n\ge a
$$ 
where 
$B$ refers to
the incomplete beta function
$$
B\left( \frac{\theta^\gamma}{\theta^\gamma + a^\gamma};
\frac{\beta_2-1}\gamma,\frac{1-\beta_1}\gamma
\right)
$$
and
the extra parameter $\gamma >0$ controls the sharpness of the transition
\cite{footnote_marc2}.
The limit $\gamma\rightarrow \infty$ recovers the (infinitely sharp) double power law
and
the limit $a=\beta_1=0$ with $\gamma=1$ leads to the Pareto distribution.
In general,
one could fix $\gamma=1$ to reduce the number of free parameters, but this seems a rather arbitrary decision.
{An even more convenient option would be to consider the discrete version of this distribution (as the data are discrete).
These extensions are left for future research.}}

A relevant issue in our analysis 
is the construction of the harmonic codewords from the scores. 
The discretization procedure 
looks for the presence (1) or absence (0) of each pitch class
in every beat of the score.
This involves two somewhat arbitrary decisions:
first,
presence and absence are decided in terms of an arbitrary
threshold applied to the (nonbinary, continuous) chromas;
second,
it is taken for granted that the fundamental time unit is the beat.
We have tested the robustness of our results against the change
in these arbitrary parameters, 
repeating the process for different discretization thresholds
and different selections of the elementary time unit.
It is clear that the increase of the time unit 
(e.g., from one beat to two beats) leads to a reduction in 
the number of tokens $L$
and subsequently in the values of the frequencies $n$.
Obviously, scale parameters (such as $\theta$ and $e^\mu$)
are strongly affected under such a change;
however, a rescaled plot such as the one in
Fig. \ref{fig_rosso}(b)
shows that the shape of the distributions is robust
and remains nearly the same,
also when the threshold is changed.
As the distinction between lognormals and power laws depends
on the shape and not on the scale of the distributions, 
our conclusions regarding the lack of universality 
and the poor fulfilment of Zipf's law in music do not change
(in a previous study utilizing the same discretization method, 
we also found the definition of the codeword threshold and the temporal unit rather irrelevant given a reasonable range \cite{Serra_Corral_richness}).

In summary, we find that the usage of harmonic vocabulary 
in classical composers may seem universal-like at a qualitative level
(see Fig. \ref{fig_rosso}(b)),
but this paradigm fails when one approaches the issue in a quantitative statistical way.
Not only universal parameters to describe the distributions do not exist, but different distributions (lognormal and power laws) 
fit better different composers.
In particular, the Zipf picture (a power-law tail with exponent in the approximate range $1.8$--$2.2$) only applies to a reduced subset of composers 
{(9 best fits, out of 76 composers, 
and 22 good fits, out of 69 power law and double power-law good fits
and 76 composers, 
see Table \ref{tab:summary})}.
Although some degree of universality has been claimed in complex systems in analogy to statistical physics \cite{Stanley_rmp},
detailed analyses {in some particular systems} 
have shown a diversity of parameters and distributions 
for some particular systems
\cite{Hantson_Pueyo16,Corral_Gonzalez}.

{Our work can be put into the wider context of quantifying the universality of scaling laws. 
This is directly related to the use of proper statistical methodologies
in complex-systems science, 
where some controversies have arisen in recent years.
For example, the application of the ideas of allometric scaling to urban science 
\cite{Bettencourt_West} has raised important concerns 
\cite{Arcaute_scaling,Leitao} (see also Ref. \cite{Ballesteros} for a revision of the original problem). 
Methods of fitting power-law distributions have been criticized  
\cite{Clauset}
and re-criticized 
\cite{Corral_nuclear,Barabasi_criticism,Voitalov_krioukov,Corral_epidemics_pre}.
The important role played by statistical dependence when fitting has been pointed out
in Ref. \cite{Gerlach_Altmann_prl}.
In general, one important lesson that emerges from the study of complex systems is that these have to be characterized in probabilistic terms, so a precise description of their stochastic or probabilistic properties, together with adequate statistical tools, becomes mandatory.
}


%

\section{Acknowledgements}

Josep Llu\'{\i}s Arcos
provided support to start this collaboration;
Juan Camacho contributed with discussions about statistical physics;
V. Navas-Portella facilitated his R-code for fitting;
I. Moreno-S\'anchez participated in a preliminary stage of this research.
Support from projects
FIS2015-71851-P and
PGC-FIS2018-099629-B-I00
from Spanish MINECO and MICINN,
as well as from
CEX2020-001084-M from
the Spanish State Research Agency, 
through the Severo Ochoa and Mar\'{\i}a de Maeztu Program for Centers and Units of Excellence in R\&D,
is acknowledged.
M. S.-P.'s participation has been possible thanks to the Internship Program of the Centre de Recerca Matem\`atica
and Treball Campus of the UAB.

\section{Author Contributions}

J. S. and A. C. discussed the original idea, 
M. S.-P. wrote the code and did a first analysis of the data, 
A. C. did a second analysis, 
J. S. and A. C. supervised the results, all authors interpreted the results. 
A.C. wrote the first draft of the manuscript and all authors revised the manuscript. 
All authors read and approved the final manuscript.

\section{Competing Interests}
The authors declare no competing interests.

\newpage
\section{Supplementary Information}
\title{
Supplementary Information:\\
Lognormals, Power Laws and Double Power Laws
in the Distribution of Frequencies
of Harmonic Codewords from 
Classical Music
} 
\author{Marc Serra-Peralta}
\affiliation{%
Centre de Recerca Matem\`atica,
Edifici C, Campus Bellaterra,
E-08193 Barcelona, Spain
}
\affiliation{Departament de Física,
Facultat de Ci\`encies,
Universitat Aut\`onoma de Barcelona,
E-08193 Barcelona, Spain
}
\author{Joan Serr\`a}
\affiliation{Dolby Laboratories, E-08018 Barcelona, Spain}
\author{\'Alvaro Corral}
\affiliation{%
Centre de Recerca Matem\`atica,
Edifici C, Campus Bellaterra,
E-08193 Barcelona, Spain
}\affiliation{Departament de Matem\`atiques,
Facultat de Ci\`encies,
Universitat Aut\`onoma de Barcelona,
E-08193 Barcelona, Spain}
\affiliation{Complexity Science Hub Vienna,
Josefst\"adter Stra$\beta$e 39,
1080 Vienna,
Austria
}

\maketitle

\section{Data Processing and Fitting}

\subsection{Data Cleaning and Detection of Repeated Pieces}

As mentioned before, very short pieces are removed, and for us this is given by a number of beats below or equal to 15.
In order to identify files corresponding to repeated pieces, we check if two or more files 
yield the same values of $L$ and $V$. 
As this is not enough to characterize a repetition, we do the same analysis
for bigrams, trigrams, and quadrigrams
(a bigram is a concatenation of two codewords, a trigram of three, etc.),
and the pieces having the same values of $L$ and $V$ from unigrams (our standard codewords)
to quadrigrams are visually inspected, and the repeated ones removed.
The number of repetitions was found to be 212.
The resulting final values of $L$ and $V$ for the global corpus
(76 composers with 9327 pieces) are $L=5,088,442$
and $V=4085$.



\subsection{Harmonic Codewords}

The elementary entities 
of which each musical piece is composed 
are defined as in Ref.~\cite{Serra_Corral_richness} 
(in a similar way as it was done before in Ref. \cite{Serra_scirep}).
A summary of the procedure to obtain them follows:

\begin{enumerate}[(i)]
\item Starting with a MIDI file, corresponding to the score of a piece,
this is converted
by means of the {\tt midi2abc} program \cite{midi2abc}
into a standard readable text file,
containing the time occurrence, duration, and pitch of each note.
\item All pitches are transformed into pitch classes, 
which removes the distinction between different octaves.
This leads to 12 pitch classes, $C$, $C\#$, etc., up to $A\#$ and $B$.
\item The piece is divided into time intervals
corresponding to beats, as the beat is the fundamental time scale in music.
Silent time intervals
at the beginning and at the end of the piece are removed.
\item 
For each time interval,
the sum of the duration of all the notes 
of a given pitch class
are assigned to one of the
12 components of a 12-dimensional vector (called chroma),
with each component associated to a pitch class, 
$(C,C\#,\dots G\#, A,A\#,B)$.
Notes that occupy more than one time interval
are split {proportionally between the intervals}.
The duration is measured in units of the time interval
(i.e., in units of beats).
\item The vectors are discretized,
with components below a fixed threshold (0.1 by default) reset to zero
and components above the threshold reassigned to one.
These ``discretized chromas'' yield the harmonic codewords
or types
that constitute 
the harmonic building blocks of the musical pieces. 
Note that there are $2^{12}=4096$ possible types.
So, at this coarse-grained level,
each piece is represented as a 12-dimensional time series of binary elements.
\item {Each piece 
is transposed;} 
major pieces to $C$ major and minor pieces to $A$ minor.
This constitutes a shift of the pitches by a constant amount 
(or a rescaling of the wavelengths, in physical terms), 
in order that the tonic note becomes pitch class $C$ 
(for major keys) or $A$ (minor keys).
Thus, pitch classes turn to represent in fact tonal functions.
See Ref. \cite{Serra_Corral_richness},
{which also details the method used for the determination of the key.}
\item All pieces corresponding to the same composer are aggregated.
This is done at the level of the time series of (transposed) discretized chromas.
The reason for aggregation 
is that many pieces are too short to provide a meaningful statistical analysis.
The resulting aggregation yields 76 datasets, one for each composer.
{An additional dataset, further aggregating the 76 individual datasets, is considered as representative of ``classical music'' as a whole.
We refer to it as the global dataset.}
\end{enumerate}

Notice that step vi (transposition) would be irrelevant if we calculated the distribution of codeword counts for individual pieces, but it becomes very important as we aggregate the pieces, 
because they usually come in different keys
(transposition makes the key to be the same, for pieces in major and minor keys, separately).

\subsection{Fitting Method}

For each dataset,
the absolute frequencies $n$ of each of the $V$ types 
provides the values of the random variable for which the statistical analysis is performed
(other approaches use the so-called rank as the random variable \cite{Altmann_Gerlach},
but there are good arguments in favor of using $n$ \cite{Corral_Cancho}).
The set of values of $n$ allow one to apply maximum-likelihood estimation
(once a fitting distribution is provided), 
as well as the Kolmogorov-Smirnov goodness-of-fit test
(the latter through the construction of the estimated cumulative distribution function).
In principle, 
a fit is accepted (in the sense that it is not rejected)
if the resulting $p-$value of the Kolmogorov-Smirnov test is greater than 0.20
(note that this is much stricter than the usual 0.05 significance level).
$p-$values are calculated from 100 Monte-Carlo simulations of the distribution resulting from the fit.

Fits are not performed for all $n\ge 1$, 
but above the lower cut-off $a$, greater than one in practice.
The reason is that it is nearly impossible that a simple probability distribution (with few parameters) fits well the smallest frequencies, for which the statistics is very high
(which is a common characteristics of Zipfian systems \cite{Corral_Cancho});
we systematically obtain $p=0$ if $a=1$.
Thus, fits are repeated for different values of the lower cut-off,
which are swept uniformly in log-scale 
(with 20 values per order of magnitude).
In the case of the truncated power-law, there is also an upper cut-off $b$ 
in the definition of the distribution, and, in the same way, 
different values of the upper cut-off are analyzed.

Among all non-rejectable fits 
(all with $p>0.20$, for different values of the cut-offs), 
the procedure we use selects
the one that includes more types.
If the resulting value of the lower cut-off $a$ is larger than 32, 
the fit is rejected (despite $p>0.20$),
due to the fact that more than one and a half orders of magnitude in frequencies (from $n=1$ to $10^{3/2}\simeq 32$) 
would be excluded from the fit.
Thus, $p>0.20$ and $a\le 32$ is what we consider a ``good fit''.
{Note that
when we quantify the span of
the fitting range of a distribution 
we refer to the logarithmic span of the fitting range, 
$n_{max}/a$, 
with $n_{max}$ the maximum value of $n$ observed in the data.}
%
%
An important thing about the fitting method
is that it is fully automatic and does not involve subjective judgements 
(so as to decide when the power-law regimes start).
More details 
are given in Refs. \cite{Corral_Deluca,Corral_Gonzalez}.

\subsection{Model Selection}


 
When
more than one distribution 
fits the data, 
we need to decide which is the best fit, in order to select it 
as the one best explaining the data.
{For model comparison, it is important to correct 
for the different number of free parameters that each distribution may have
(the more parameters, the better the fit, in principle, but the improvement has to be significant).
For that purpose, it is common to use the Akaike and the Bayesian information criteria, or a likelihood-ratio test \cite{Clauset}.

The problem we deal with, however, has an extra ingredient that makes the situation more complicated:
each distribution fits a different subset of the data, in general,
given by different values of the lower cut-off $a$
for each distribution
{($a_\text{pl}$, $a_\text{dpl}$ and $a_\text{ln}$)}; 
i.e., only values of the random variable fulfilling $n\ge a$ are fitted
(the criteria mentioned in the previous paragraph 
compare fits acting over fixed data).}
To overcome this difficulty,
we use the simple criterion of choosing the distribution that fits a larger number of data points
(i.e., more types), 
which is equivalent to select the one with the smaller lower cut-off $a$,
independently of the number of parameters.
This is what defines for us the ``best fit,''
and 
in this way, we give absolute priority to fitting the largest fraction of the dataset.

{The reason for our choice can be justified as follows.
If a distribution fits the data for $n\ge a$, then, the remainder $a-1$ empirical points (outside the fit) can be considered to be fitted using the $a-1$ values
of the empirical probabilities $f(n)$, from $f(1)$ to $f(a-1)$.
Thus, if the difference in $a-$values between two distributions is relatively large, 
as it is the usual case, the difference in the number of ``total parameters''
(including the empirical probabilities below $a$) is large as well
and the distribution with smaller $a$ has a much smaller number of ``total parameters.''
Note that, in any case, 
we are comparing distributions that already provide good fits 
($p>0.20$).
Moreover, we find more natural to prefer a distribution given by a simple formula,
than one containing empirical values of probabilities as free parameters.
}
In the case of two distributions yielding the same minimum value of the lower cut-off (tie), 
the distribution with a smaller number of parameters is selected as the best fit
{(we only find this situation in 7 cases).}


\begin{thebibliography}{10}
\bibitem{footnote_marc}
In the fitting of the truncated power law, the method we use
  \cite{Corral_Deluca,Corral_Gonzalez} considers the possibility of
  $b>n_\text{max}$ (the truncation point could be above the maximum value of
  the random variable). for the case we are interested in (the truncated power
  law giving one of the regimes of the double power law), this does not apply,
  and therefore we have ignored this fact.
\end{thebibliography}

\end{document}